\documentclass[journal]{IEEEtran}
\usepackage{amsmath}
\usepackage{amssymb}
\usepackage{amsthm}
\usepackage{bm}
\usepackage{pgfplots}
\usepackage{pifont}
\usepackage[nofillcomment]{algorithm2e}
\RestyleAlgo{ruled}
\usepackage{amsfonts}
\usepackage{bbm}
\usepackage{mathtools}
\usepackage{setspace}
\usepackage{multicol}

\usepackage{float}
\usepackage{cite}
\newtheorem{Lemma}{Lemma}
\newtheorem{Theorem}{Theorem}
\newtheorem{Corollary}{Corollary}
\newtheorem{Proposition}{Proposition}
\newtheorem{Remark}{Remark}
\newtheorem{Definition}{Definition}

\definecolor{mycolor1}{rgb}{0.00000,0.44700,0.74100}
\definecolor{mycolor2}{rgb}{0.85000,0.32500,0.09800}
\definecolor{mycolor3}{rgb}{0.4660, 0.6740, 0.1880}

\usepackage{color}

\newcommand{\RQ}{\mathcal{R}_q}
\newcommand{\alsize}{Q}
\newcommand{\coefffailprob}{\textup{Pr}(\mathcal{E})}
\newcommand{\decfailprob}{\text{DFR}}
\newcommand{\compnoisev}{c_{N_v}}
\newcommand{\compnoiseu}{c_{N_u}}

\newcommand{\encmap}{\textsf{Encode/Map}}
\newcommand{\demapdec}{\textsf{Demap/Decode}}
\DeclareMathOperator{\supp}{supp}
\newcommand{\decomp}{\textsf{decomp}_{q}}
\newcommand{\comp}{\textsf{comp}_{q}}

\usetikzlibrary{external}
\usepackage{stackengine} 
\newcommand\oast{\stackMath\mathbin{\stackinset{c}{0ex}{c}{0ex}{\ast}{\bigcirc}}}

\newcommand{\remove}[1]{}

\makeatletter
\newcommand{\removelatexerror}{\let\@latex@error\@gobble}
\makeatother

\newcommand{\myalgorithm}[1]{%
\begingroup
\removelatexerror
\begin{algorithm*}[H]
	#1
\end{algorithm*}
\endgroup}

\begin{document}
	\title{Information- and Coding-Theoretic Analysis of the RLWE/MLWE Channel}

	\author{Georg Maringer,~\IEEEmembership{Student Member,~IEEE,}
	Sven Puchinger,~\IEEEmembership{Member,~IEEE,}
	Antonia Wachter-Zeh,~\IEEEmembership{Senior Member,~IEEE}
	\thanks{Parts of this work have been presented at the IEEE Information Theory Workshop (ITW) 2020 \cite{maringer2020higher}.}
	\thanks{G. Maringer and A. Wachter-Zeh are with the Department of Electrical and Computer Engineering at the Technical University of Munich. S. Puchinger was with the Department of Applied Mathematics and Computer Science at the Technical University of Denmark. He is now with Hensoldt Sensors GmbH. Emails: \{georg.maringer, antonia.wachter-zeh\}@tum.de, mail@svenpuchinger.de}
	\thanks{G. Maringer's work was supported by the German Research Foundation (Deutsche Forschungsgemeinschaft, DFG) under Grant No. WA3907/4-1. S.~Puchinger received funding from the European Union's Horizon 2020 research and innovation program under the Marie Sklodowska-Curie grant agreement no.~713683.
	A.~Wachter-Zeh and S.~Puchinger were supported by the European Research Council (ERC) under the European Union’s Horizon 2020 research and innovation programme (grant agreement no.~801434)}}
	
	\maketitle
	\begin{abstract}
		Several cryptosystems based on the \emph{Ring Learning with Errors} (RLWE) problem have been proposed within the NIST post-quantum cryptography standardization process, e.g., NewHope. Furthermore, there are systems like Kyber which are based on the closely related MLWE assumption. Both previously mentioned schemes result in a non-zero decryption failure rate (DFR).
		The combination of encryption and decryption for these kinds of algorithms can be interpreted as data transmission over a noisy channel.
		To the best of our knowledge this paper is the first work that analyzes the capacity of this channel. We show how to modify the encryption schemes such that the input alphabets of the corresponding channels are increased.
		In particular, we present lower bounds on their capacities which show that the transmission rate can be significantly increased compared to standard proposals in the literature.
		Furthermore, under the common assumption of stochastically independent coefficient failures, we give lower bounds on achievable rates based on both the Gilbert-Varshamov bound and concrete code constructions using BCH codes.
	By means of our constructions, we can either increase the total bitrate (by a factor of $1.84$ for Kyber and by factor of $7$ for NewHope) while guaranteeing the same DFR or for the same bitrate, we can significantly reduce the DFR for all schemes considered in this work (e.g., for NewHope from $2^{-216}$ to $2^{-12769}$).
	\end{abstract}

	\begin{IEEEkeywords}
		Ring/Module LWE, RLWE/MLWE channel, channel capacity, error correcting codes
	\end{IEEEkeywords}
	
	\vspace{-0.5em}
	\section{Introduction}
	The security of most currently deployed asymmetric encryption schemes as well as digital signatures is based on the hardness of integer factorization or the discrete logarithm problem. In 1999, Shor developed quantum algorithms that are able to solve both of these problems in polynomial time with respect to the size of the integer to factorize or the size of the group over which the discrete logarithm problem is defined~\cite{shor1999polynomial}.
	Quantum computers with a sufficient amount of qubits to actually break schemes like RSA do not exist yet. However, driven by companies like Google and IBM significant progress has been made recently.
	Hence, it is essential to develop \emph{post-quantum} (PQ)-secure cryptographic schemes due to the requirement of long-term security for devices that are hard to update (e.g., satellites). Furthermore, investigation of cryptographic schemes and the development of hardware and software implementations is a challenging task which requires time and effort of the cryptographic community.
	
	The security of several encryption and signature schemes that are considered in the NIST-PQC standardization process \cite{NIST_PQC} is based on the hardness of certain problems on lattices. We refer to this family of primitives as \emph{lattice-based schemes}. Several of these schemes are based on the \emph{Learning with Errors} (LWE) problem which was shown to be reducible to the decisional version of the \emph{Shortest Vector Problem} (SVP) and the \emph{Shortest Independent Vectors Problem} (SIVP) on lattices~\cite{regev2009lattices}. The security of the schemes examined in this work are based on the closely related \emph{Ring Learning with Errors} (RLWE) and the \emph{Module Learning with Errors} (MLWE) problems. The former can be reduced to the (approximate) SIVP problem in a subclass of lattices, so-called ideal lattices \cite{lyubashevsky2013ideal} and the latter one to (approximate) SIVP problem on module lattices. Cryptographic schemes based on these problems result in a smaller key size compared to those based on the LWE problem. Computations in RLWE/MLWE-based schemes can be implemented very efficiently (e.g., by using the \emph{number theoretic transform} (NTT) for the polynomial multiplications).

	In~\cite{lee2019modification} it was suggested to view the LWE-based cryptographic scheme Frodo~\cite{bos2016frodo} as a digital communication system.
	Exchanging messages between two parties in a secure manner using RLWE/MLWE-based algorithms can also be considered as data transmission over a noisy channel, in the following referred to as the \emph{RLWE/MLWE channel}.
	Consequently, we can find the channel capacity of this cryptographic channel by using Shannon's noisy channel capacity theorem presented in \cite{shannon1948mathematical}. To our knowledge, this paper is the first that analyzes the capacity and other information-theoretic properties of the {RLWE/MLWE channel}.
	In \cite{rajagopalan2018wiretap} polar coded LWE-based symmetric key encryption schemes as well as wiretap coded LWE-based encryption have been investigated, while in this work we analyze LWE/RLWE/MLWE-based public key encryption schemes.
	
	It is possible to choose the parameter sets of LWE/RLWE/MLWE based public key encryption schemes such that the decryption of the ciphertext never fails if the recipient knows the private key. Although this property is desirable, there are practical reasons why a non-zero (but very small) \emph{decryption failure rate} (DFR) is permitted by several algorithms (e.g., $2^{-174}$ for Kyber). It significantly reduces key sizes, the size of the ciphertext (for the same message length) and the complexity of the encryption and decryption algorithms.
	A low decryption failure rate is not only essential since retransmissions cost data rate but they also provide information for an attacker that tries to break the cryptosystem \cite{fluhrer2016cryptanalysis}.
	A possible measure to decrease the DFR is to use suitable \emph{error-correcting codes} (ECC).
	For an analysis on the impact of ECCs in NewHope Simple \cite{alkim2016newhope} see \cite{fritzmann2018analysis}. In their work the effect of using a BCH code, an LDPC code and their concatenation is analyzed. However, only the influence of one specific BCH code is analyzed whereas in this work we optimize the BCH parameters with respect to different alphabet sizes.
	The analysis presented in this work provides a framework and can be applied to various LWE/RLWE/MLWE based schemes. For the analysis we chose to consider Kyber and NewHope within the main part of this paper. The reason for these choices is that Kyber is still a main candidate within the Round 3 of the NIST PQC competition whereas NewHope has already been practically examined by Google as a candidate to achieve post quantum secure communication \cite{braithwaite2016}. Furthermore, we provide results for Frodo and LAC in the Appendix to show that it the framework can easily be applied to other relevant schemes. The two other lattice-based schemes NTRU and Saber are both still part in the Round 3 of the NIST PQC. However, our framework can only be applied to schemes that with non-zero DFR. Therefore, we did not consider NTRU as the parameter choices for this scheme are designed such that decryption failures are impossible. Saber is a Learning with Rounding (LWR)-based scheme and the noise creation is therefore significantly different from the schemes considered in this work.

	In Section~\ref{sec:preliminaries}, we introduce basic notation, lattices, some coding fundamentals and define ciphertext compression and decompression functions.
	Section~\ref{sec:RLWE_cryptography} deals with the basics of RLWE/MLWE-based cryptography and lattice-based cryptography in general. In Section~\ref{sec:RLWE_channel}, we show how to connect RLWE/MLWE-based cryptosystems to communication theory in the Shannon sense. The consequences of these results naturally leads to the information-theoretic analysis presented in Section~\ref{sec:information_theoretic_LWE}. In this section we also show how to bound the decryption failure rate of RLWE/MLWE-based schemes under the assumption of stochastically independent coefficient failures. Section~\ref{sec:semi_constructive_analysis} deals with maximizing the achievable rates of the considered schemes and with the minimization of the decryption failure for fixed minimal bitrates using ECCs. Finally in Section~\ref{sec:conclusion} we sum up the results and conclude the paper.
	
	\section{Preliminaries}\label{sec:preliminaries}
	\subsection{Notation}
	Throughout this work, polynomials are either denoted as lowercase letters or lowercase letters with the indeterminate in brackets, e.g. polynomial $a$ or $a(x)$, respectively. For a polynomial $a$, the $i$-th coefficient is denoted as $a_i$ unless otherwise mentioned. Vectors are denoted by lowercase letters in bold font, e.g. $\bm{v}$ and its $i$-th component as $v_i$. We denote matrices with polynomial components by bold uppercase letters, e.g. $\bm{A}$ and the polynomial in the $i$-th row and  $j$-th column by $A_{ij}$.
	
	Sampling an element $b$ from a distribution $\chi$ is denoted by $b \xleftarrow[]{\text{\$}} \chi$ and sampling uniformly from a set $\mathcal{S}$ is denoted by $b \xleftarrow[]{\text{\$}} \mathcal{S}$. Independent sampling of every coefficient of a polynomial $a \in \RQ$ according to a distribution $\chi$ is denoted by $a \xleftarrow[]{\text{\$}} \chi(\RQ)$ and independent sampling of a vector $\bm{v} \in \RQ^l$ according to $\chi$ is denoted by $\bm{v} \xleftarrow[]{\text{\$}} \chi(\RQ^l)$. We denote the binomial distribution by $\mathcal{B}(i,n,p)$, where $i$ specifies the number of successes, $n$ the number of trials and $p$ the success probability. We define the magnitude of an element in $\mathbb{Z}_q$ by the magnitude of its representation in the interval $[-q/2,q/2]$.
	The rounding operator is denoted by $\lceil . \rfloor$, where in particular $\lceil x.5 \rfloor=x+1$.

	Let $P_X*P_Y$ denote the convolution of two probability mass functions $P_X$ and $P_Y$ and let the $n$-fold convolution of $P_X$ with itself be $\oast_n(P_X) := P_X * P_X * \dots * P_X$, in particular $\oast_1(P_X) = P_X * P_X$ and $\oast_0(P_X) = P_X$.
	
	\begin{Definition}[Centered binomial distribution]\label{def:cent-bin-dist}
		The centered binomial distribution with parameter $k$, denoted as $\chi_k$, is defined as $\chi_k(x) := \mathcal{B}(x+k/2,k,1/2)$, where $x \in \{-k/2,-k/2+1, \dots, k/2\}$.
	\end{Definition}
	The expectation of the centered binomial distribution $\chi_k$ is $0$ and its variance is~$k/2$.
	It is possible to sample relatively efficiently from this distribution compared to, e.g., the rounded Gaussian distribution.

	\begin{Definition}[Lattice]
		A lattice $\mathcal{L}$ is defined as the set of linear combinations over the integers $\mathbb{Z}$ of a set of linearly independent vectors $\bm{b}_1, \dots, \bm{b}_n \in \mathbb{R}^m$.
		\begin{equation*}
			\mathcal{L} := \left\{ \sum_{i=1}^n \alpha_i \bm{b}_i: \alpha_i\in \mathbb{Z}, i=1,\dots,n \right\}
		\end{equation*}
	\end{Definition}

	There are several computationally hard problems defined on lattices. A detailed description of the examples shown below can be found in \cite{hoffstein2008introduction} and \cite{peikert2016decade}. The underlying problems used for the security reductions of the algorithms presented in this work are all related to lattices.
	\begin{Definition}[Negligible Function, \cite{katz2014introduction}]
		A negligible function $f$ from the natural to the non-negative real numbers in some parameter $\lambda$ satisfies that there exists a number $N$ such that for all $\lambda>N$ it holds that
			$f(\lambda) < 1/p(\lambda)$ for every positive polynomial $p$.
	\end{Definition}

	\subsection{The Ring $\RQ$}
	Let $\RQ := \mathbb{Z}_q[x]/(x^n+1)$ be the polynomial ring in $x$ of degree $n$ with coefficients in $\mathbb{Z}_q$.
	The addition of two polynomials in $\RQ$ is performed by adding the coefficients in $\mathbb{Z}_q$:
	\begin{equation*}
		a(x) + b(x) = c(x), \quad \text{where } c_k = a_k + b_k \mod q.
	\end{equation*}
	The multiplication of two polynomials in $\RQ$ is defined by
	\begin{equation*}
		a(x) \star b(x) = c(x), \ \text{where } c_k = \sum_{i=0}^k a_i b_{k-i} - \sum_{i=k+1}^{n-1} a_i b_{n-i+k},
	\end{equation*}
	for all $k =0,\dots,n-1$.
	Thus, each polynomial in $\mathbb{Z}_q[x]/(x^n+1)$ can be represented by a polynomial in $\mathbb{Z}_q[x]$ of degree $<n$.
	Frequently throughout this work, we use $ab$ as a shorthand notation for the multiplication of two polynomials $a,b \in \RQ$.

	\subsection{Linear Codes}
	The parameters of a linear (block) code $\mathcal{C}$ over $\mathbb{F}_q$ are denoted by $[n,k,d]_q$, where $n$ is the length, $k$ its dimension, and $d$ its minimum Hamming distance.
	
	The \emph{Gilbert--Varshamov} (GV) bound \cite{gilbert1952comparison,varshamov1957estimate} can be used to show that codes with certain parameters exist. It states that for any parameter set $[n,k,d]_q$ fulfilling the inequality
		
	\begin{equation}
		q^{n-k} > \sum_{i=0}^{d-2} \binom{n-1}{i} (q-1)^i,
	\end{equation}
	there exists a linear $[n,k,d]_q$ code. The bound is non-constructive, i.e., it does not give an efficient algorithm to construct such a code. For this reason, we also consider the more practical class of BCH codes \cite{bose1960class}, \cite{hocquenghem1959codes} within cryptographic schemes to reduce their decryption failure rates. Whenever we specify their minimum distance we refer to the designed minimum distance of the associated RS supercode. The actual minimum distances of a BCH code can (and is likely to be) even larger than specified. For an elaborate introduction of BCH codes and the GV bound we refer to \cite{roth_2006}.
	
	\subsection{Ciphertext compression and decompression}
	Within the algorithms used in this work coefficients of polynomials are frequently compressed to reduce the size of the generated ciphertexts. At the receiver a decompression function is applied. Since the compression is lossy the concatenation of compression and decompression only approximates its input.
	
	We denote the output of the compression function on input $z$ by $z'$. The compression function compresses each coefficient down to $d_c$ bits and is defined by 
	\begin{equation}
		z'=\comp(z,d_c) = \left\lceil \tfrac{z\cdot 2^{d_c}}{q}\right\rfloor \mod 2^{d_c}\enspace .
	\end{equation}
	We denote the output of the decompression function on input $z'$ by $z''$. The decompression function is defined by 
	\begin{equation}
		z''=\decomp(z',d_c) = \left\lceil \tfrac{z' \cdot q}{2^{d_c}} \right \rfloor \enspace .
	\end{equation}
	The inputs to $\comp$ and $\decomp$ are to be represented in the range $\{0,\dots,q-1\}$.
	We define both functions also for vectors of polynomials in $\RQ$ by applying them separately to each polynomial. The exact specification of the ciphertext compression within NewHope and Kyber can be found in \cite{alkim2016newhope} and \cite{Kyber_supp}, respectively.
	
	\section{RLWE/MLWE-based Cryptography}\label{sec:RLWE_cryptography}
	
	\subsection{RLWE and MLWE Problem}
	The \emph{Learning with Errors} (LWE) problem was introduced by Regev \cite{regev2005}. Several cryptosystems are basing their security on the hardness of LWE. The complexity of encryption and decryption of those schemes can be improved for systems basing their security on the closely related but more structured \emph{Ring Learning with Errors} (RLWE) \cite{lyubashevsky2013ideal} or Module Learning with Errors (MLWE) \cite{langlois2015worst} problems. Furthermore, the comparably large key sizes of LWE-based schemes can considerably be shrunk for RLWE/MLWE-based schemes.
	
	\begin{Definition}[RLWE problem]
		Consider a set of samples of the form
		\begin{equation}\label{eq:RLWE_samples}
			(a_i,b_i = a_i s + e_i),\ i=1,\dots,m,
		\end{equation}
		where the $a_i$ are drawn from the uniform distribution on $\RQ$ and $s$ as well as the $e_i$ are sampled from $\chi_k(\RQ)$. The decisional version of the RLWE problem is defined to be the task of distinguishing samples drawn from the distribution specified in \eqref{eq:RLWE_samples} from samples drawn from the uniform distribution on $\RQ \times \RQ$, where the problem shall be solved correctly with an advantage being non-negligible compared to random guessing, i.e. the advantage should be lower bounded by a function which is not negligible in the security parameter.
	\end{Definition}
	\begin{Definition}[MLWE problem]\label{def:MLWE}
		Consider a set of samples of the form
		\begin{equation}\label{eq:MLWE_samples}
			(\bm{a}_i,b_i=\bm{a}_i \bm{s} + e_i),\ i=1,\dots,m,
		\end{equation}
		where the $\bm{a}_i$ are drawn from the uniform distribution on $\RQ^l$, $s$ is sampled from $\chi_k(\RQ^l)$ and the $e_i$ are sampled from $\chi_k(\RQ)$. 
		The decisional MLWE problem is defined to be the task of distinguishing samples drawn from the distribution specified in \eqref{eq:MLWE_samples} from samples drawn from the uniform distribution on $\RQ^l \times \RQ$, where the problem shall be solved correctly with an advantage being non-negligible compared to random guessing, i.e. the advantage should be lower bounded by a function which is not negligible in the security parameter.
	\end{Definition}
	
	\subsection{Public Key Encryption based on MLWE/RLWE}
	We consider the scenario that Alice would like to transmit a message to Bob using a public key encryption scheme. In order to do this, Bob generates a key pair $(pk,sk)$ consisting of a public key $pk$ and a secret key $sk$. The public key is then used by Alice to encrypt a message $m$ to obtain a ciphertext $c$ which she sends to Bob. Bob then uses his secret key (private key) $sk$ and the ciphertext $c$ to obtain an estimate (remember decryption may fail with very small probability) for the message $m$.
	Each \emph{public key encryption} (PKE) scheme is composed of three functions: key generation, encryption, and decryption. The first one generates the required public and private keys, the second one is for encryption and the last one is for decryption. The basic building blocks for RLWE/MLWE-based schemes are presented in Algorithms~\ref{algorithm:key_gen_MLWE},~\ref{algorithm:encryption_MLWE}~and~\ref{algorithm:decryption_MLWE}.

	\begin{figure}[t]
	\begin{center}
	\begin{minipage}{\columnwidth}
		\myalgorithm{
			\small
			\KwIn{$n,q,k,l$}
			$\boldsymbol{A} \xleftarrow[]{\text{\$}} \RQ^{l \times l}$ \\
			$\boldsymbol{s},\boldsymbol{e} \xleftarrow[]{\text{\$}} \chi_k(\RQ^l)$\\
			$\boldsymbol{b} \leftarrow \boldsymbol{A}\boldsymbol{s}+\boldsymbol{e}$\\
			\KwResult{$pk=(\bm{A}, \bm{b})$, $sk = \boldsymbol{s}$}
			\caption{Key Generation}
			\label{algorithm:key_gen_MLWE}
		}
		\myalgorithm{
			\small
			\KwIn{$pk=(\bm{A},\bm{b})$, $m \in \mathcal{M}$, $(n,q,k,l)$, $d_u,d_v$}
			$\boldsymbol{s'},\boldsymbol{e'} \xleftarrow[]{\text{\$}} \chi_k(\RQ^l)$ \\
			$e'' \xleftarrow[]{\text{\$}} \chi_k(\RQ)$\\
			$\boldsymbol{u} \leftarrow \bm{A}^T\bm{s'} + \boldsymbol{e'}$ \\
			$v \leftarrow \bm{b}^T \bm{s'} + e'' + \encmap(m)$\\
			$\bm{u'} \leftarrow \comp(\bm{u},d_u)$\\
			$v' \leftarrow \comp(v,d_v)$\\
			\KwResult{$c= (\bm{u'}, v')$} 
			\caption{Encryption}
			\label{algorithm:encryption_MLWE}
		}
		\myalgorithm{
			\small
			\KwIn{$c= (\bm{u'}, v')$, $sk=\boldsymbol{s}$, $(n,q,k,l)$, $d_u,d_v$}
			$\bm{u''} \leftarrow \decomp(\bm{u'},d_u)$\\
			$v'' \leftarrow \decomp(v',d_v)$\\
			$\hat{m} \leftarrow \demapdec(v'' - \bm{s}^T \bm{u''})$\\
			\KwResult{$\hat{m}$}
			\caption{Decryption}
			\label{algorithm:decryption_MLWE}
		}
		\vspace{-2em}
	\end{minipage}
	\end{center}
	\end{figure}
	Within RLWE/MLWE based cryptosystems the parameter $n$ denotes the number of coefficients within the polynomials of the ring $\RQ$, $q$ denotes coefficient modulus of $\RQ$, $k$ parametrizes the error distribution (e.g., controls its variance) and $l$ specifies the dimension of the matrices and vectors used within the algorithms.
	For RLWE based schemes it holds that $l=1$. Additionally, in some RLWE-based (e.g. \cite{alkim2019newhope}) or Module-LWE based schemes (e.g. \cite{Kyber_supp}) a technique called ciphertext compression is used to reduce the size of the ciphertext.
	In comparison to schemes which are based on the Learning with Rounding problem~\cite{banerjee2012pseudorandom} (e.g., \cite{d2018saber}, \cite{bhattacharya2018round5}), ciphertext compression for RLWE or MLWE-based schemes plays only a secondary role concerning the security of the encryption schemes.
	
	\subsection{Key Generation, Encryption and Decryption}
	For the following description of key generation, encryption and decryption it is valid to consider RLWE-based scheme as MLWE-based scheme with parameter $l=1$.
	The random elements sampled in the presented algorithms are either sampled from the uniform distribution or from the error distribution.
	
	The key generation for RLWE/MLWE based schemes is shown in Algorithm~\ref{algorithm:key_gen_MLWE}. First, a matrix $\bm{A}$ is sampled uniformly from $\RQ^{l \times l}$. Then the vectors $\bm{s}$ and $\bm{e}$ are sampled from $\chi_k(\RQ^l)$ to compute $\bm{b}=\bm{A}\bm{s}+\bm{e}$. The public key is defined to be $pk=(\bm{A},\bm{b})$ and the private key is defined to be $sk=\bm{s}$.
	
	We denote the set of possible messages by $\mathcal{M}$.
	The encryption procedure of RLWE/MLWE based schemes is shown in Algorithm~\ref{algorithm:encryption_MLWE}. It involves apart from sampling polynomials according to predefined distributions and simple algebraic operations only ciphertext compression and the encoding and mapping of the message $m \in \mathcal{M}$ via the function $\encmap$. To encrypt $m$, we first need to transform it into a polynomial in $\RQ$. We have a certain flexibility in the choice of the encoding and decoding functions which can be used to reduce the overall decryption failure probability of the scheme. Within the encoding step several algorithms utilize error-correcting codes (ECCs) to reach the required DFR of the scheme, thereby achieving the desired security level. For example BCH codes are deployed in LAC \cite{LAC_supp} and NewHope \cite{alkim2019newhope} uses a repetition code of length $4$. Thus, a message that can be represented by $k$ bits is encoded by an ECC of length at most $n$, with $n$ being the number of coefficients of a polynomial in $\RQ$. Commonly the mapper takes the codeword and converts it into a polynomial in $\RQ$ by multiplying each bit of the codeword by $\left \lfloor q/2 \right \rfloor$ using the resulting sequence as the coefficients of the polynomial in sequential order.
	
	The resulting ciphertext $c$ consists of the tuple $(\bm{u'},v') \in \RQ^l \times \RQ$ and is computed according to Algorithm~\ref{algorithm:encryption_MLWE}. The steps to obtain them only involve sampling from $\chi_k$ and some finite field arithmetic in $\RQ$.
	The decryption of RLWE/MLWE based schemes is depicted in Algorithm~\ref{algorithm:decryption_MLWE}. Bob first decompresses the ciphertext using $\decomp$ and then computes 
	\begin{align}\label{eq:decoding}
	v'' - \bm{s}^T \bm{u''} &= v + \compnoisev - \bm{s}^T(\bm{u} + \bm{\compnoiseu}) \nonumber \\ 
	&= \bm{e}^T\bm{s'} + e''- \bm{s}^T(\bm{e'} + \bm{\compnoiseu}) \nonumber \\
	&\quad + \compnoisev + \encmap(m)
	\end{align}
	using his private key $\bm{s}$. We define the compression noise terms $\bm{\compnoiseu} := \bm{u''}-\bm{u}$ and similarly $\compnoisev := v''-v$.
	We split the result in equation~\eqref{eq:decoding} into two components:  $\encmap(m)$ and $\bm{e}^T\bm{s'} + e''- \bm{s}^T(\bm{e'} + \bm{\compnoiseu})+\compnoisev$, where the latter can be interpreted as a noise term composed of terms sampled from the error distribution and compression noise terms.
	
	Since all polynomials occurring within the noise are either caused by ciphertext compression or sampled  from $\chi_k$, it is likely that the coefficients of the noise are small in magnitude as long as the compression is not too strong. In order to decrypt the ciphertext, the quantity $v''-\bm{s}^T\bm{u''}$ is computed and used as the demapper's input. Arguably the simplest demapping strategy was chosen for LAC. For this scheme the demapper examines whether the $i$-th coefficient $(v''-\bm{s}^T\bm{u''})_i$ is closer to $0$ or to $\left \lfloor q/2 \right \rfloor$ modulo $q$. If the demapper's input is closer to $\left \lfloor q/2 \right \rfloor$ for the respective coefficient then the demapper outputs $1$ for the respective index $i$ in the binary output vector $\bm{d} = (d_1,\dots,d_n) \in \mathbb{Z}_2^n$, otherwise it outputs $0$.
	\begin{equation*}
	d_i = \begin{cases}
	0 \quad \text{if }|(v''-\bm{s}^T\bm{u''})_i| \leq \left\lfloor \frac{q}{4} \right \rfloor \\
	1 \quad \text{otherwise}
	\end{cases}
	\end{equation*}
	In that sense we have defined a hard decision strategy.
	\begin{Definition}
		We refer the event that $d_i \neq (\textsf{Encode}(m))_i$ as a coefficient failure in the decoding procedure. We denote it by $\mathcal{E}$ and use indices in case we specify the respective coefficient.
	\end{Definition}
	To reach the required decryption failure rate the vector $\bm{d}$ is put into the decoder afterwards (e.g. BCH decoder for LAC) which outputs an estimate of the message $\hat{m}$. Even though for other schemes soft information (e.g. NewHope) is utilized within the demapping/decoding steps, all procedures have in common that they inherently use the fact that the coefficients of the noise are small in magnitude with high probability.
	
	\begin{Definition}
		Let $m$ be the message to be transmitted from the sender to its intended recipient and let $\hat{m}$ be the output of the decoder at the receiver side. We define a decryption failure to be the event that $\hat{m} \neq m$ and denote the probability of this event as the \textbf{decryption failure rate (DFR)} of the scheme.
	\end{Definition}

	\begin{Remark}
		To reduce the size of the public key, it is common to construct the matrix $\bm{A}$ with a pseudo-random number generator (PRNG) using a seed obtained from a true random number generator. If the PRNG is cryptographically secure, it is computationally hard to distinguish the resulting matrix $\bm{A}$ from a uniform sample on $\RQ^{l\times l}$. Within NewHope \cite{alkim2019newhope} and Kyber \cite{Kyber_supp} SHAKE128 is used as a PRNG to generate $\bm{A}$.
	\end{Remark}

	\subsection{Analysis of the noise}
	
	\begin{Lemma}\label{lem:indistinguishability_ciphertext}
		The distribution of the ciphertext components before compression $(\bm{u},v) = (\bm{A}^T \bm{s'} + \bm{e'},\bm{s}^T \bm{A}^T \bm{s'} + \bm{e}^T \bm{s'} + e'' + \encmap(m))$ cannot be distinguished from the uniform distribution on $\RQ^l \times \RQ$ if the decisional MLWE problem is hard for the respective parameter set $(n,q,k,l)$. A similar statement holds for RLWE based schemes if the decisional RLWE problem is hard for the parameter set $(n,q,k)$.
	\end{Lemma}
	\begin{IEEEproof}
		The proof of this statement for RLWE/MLWE-based schemes is similar to the security proof in the binary case in \cite{lindner2011better}. We recapitulate it here for the sake of completeness.
		The result for the RLWE case follows by setting $l=1$.
		
		Writing the ciphertext tuple before ciphertext compression $\tilde{c}$ as a column vector we obtain
		\begin{equation*}
		\tilde{c} = \begin{pmatrix}
		\bm{u}\\
		v
		\end{pmatrix} = \begin{pmatrix}
		\bm{A}\\
		\bm{b}^T
		\end{pmatrix}\bm{s'}+\begin{pmatrix}
		\bm{e'}\\
		e''
		\end{pmatrix}+\begin{pmatrix}
		\bm{0}\\
		\encmap(m)
		\end{pmatrix} \enspace .
		\end{equation*}
		
		Due to the MLWE assumption $(\bm{A},\bm{b})$ cannot be distinguished from a uniform sample on $\RQ^{l \times l} \times \RQ^l$. Therefore, by Definition~\ref{def:MLWE} $\bm{b}^T \bm{s'} + e''$ can just be considered to be an additional sample in \eqref{eq:MLWE_samples}. Thus, the ciphertext component $v$ is indistinguishable from a uniformly distributed element in $\RQ$ by the MLWE assumption. Indistinguishability of $\bm{u}$ holds because $\bm{u}$ is by definition an MLWE sample.
	\end{IEEEproof}

	We consider the preimages for the output of the concatenation of compression and decompression function of some input variable $z$, i.e. 
	\begin{equation}
	\mathcal{Z}_j:=\{z \in \mathbb{Z}_q: \decomp(\comp(z,d_z),d_z)=j\} \enspace .
	\end{equation}
	The sets $\mathcal{Z}_j$ partition the set of possible inputs of the compression function $[0,q-1]$ into disjoint sets, which are determined by the target bitlength of the compression $d_z$.
	
	\begin{Corollary}\label{cor:compression_noise_uniform}
		Let $\bm{u''}=\decomp(\comp(\bm{u},d_u),d_u)$ and $v'' = \decomp(\comp(v,d_v),d_v)$ be the output of the concatenation of ciphertext compression and decompression for the ciphertext components $\bm{u}$ and $v$, respectively.
		
		Then it holds that the problem of distinguishing $Pr(\bm{u}|\bm{u''})$ and $Pr(v_i|v''_i)$ from the uniform distributions on the preimages of $\decomp(\comp(\bm{u},d_u),d_u)$ and $\decomp(\comp(v_i,d_v),d_v)$, respectively, can be reduced to the hardness of the decisional MLWE problem for $(n,q,k,l)$.
		
		Let $z''=\decomp(\comp(z,d_z),d_z)$ be the output of the concatenation of ciphertext compression and decompression for an input $z \in \mathbb{Z}_q$.
		Then it holds that $Pr(z|z'') = 1/|\mathcal{Z}_j|$.
	\end{Corollary}
	\begin{IEEEproof}
		According to Lemma~\ref{lem:indistinguishability_ciphertext} all values for $v_i$ can be considered equiprobable. Let the set of preimages of $\decomp(\comp(v_i,d_v),d_v)$ be denoted by $\mathcal{V}_{i,j}$. Each element $v_i \in \mathcal{V}_{i,j}$ leads per definition to the same output after applying the concatenation of compression and decompression to it. Therefore, it holds for $v_i \in \mathcal{V}_{i,j}$ that $Pr(v_i|v_i'')$ is equal to the uniform distribution on the set $\mathcal{V}_{i,j}$. Virtually the same argument holds for $Pr(\bm{u},\bm{u''})$.
	\end{IEEEproof}
	
	Since we have a discrete setting the interval $[0,q-1]$ cannot be subdivided into intervals of equal integer length ($q$ is in fact prime for both NewHope and Kyber), e.g., every $v_i$ is in some unique $\mathcal{V}_{i,j}$ and to each $\mathcal{V}_{i,j}$ there is an interval $\mathcal{A}_j$ associated such that $v_i=j+a$ with $a \in \mathcal{A}_j$.
	
	Recall that the distributions of $\bm{u}$ and $v$ cannot be distinguished from the uniform distributions on $\RQ^l$ and $\RQ$, respectively, according to Lemma~\ref{lem:indistinguishability_ciphertext}. Hence, the distributions of the compression noise terms can be computed coefficient-wise. The resulting distribution can be numerically computed by creating a histogram of the compression noise for the procedure of using each $z \in \{0,\dots,q-1\}$ once at the input of the concatenation compression and decompression and dividing the resulting vector by $q$.
	
	To compute the distribution of the noise we use the fact that the distribution of the sum of independent variables can be computed by convolving their respective distributions. Thus, we show in the following that that the terms $\bm{e}^T\bm{s'}$, $\bm{s}^T(\bm{e'}, \bm{\compnoiseu})$ and $\compnoisev$ are stochastically independent with overwhelming probability.
	
	\begin{Lemma}\label{lem:independence_noise}
		The set of terms $\bm{e}^T\bm{s'}$, $\bm{s}^T(\bm{e'}, \bm{\compnoiseu})$ and $\compnoisev$ is stochastically independent if there exists a triple of indices $(i,j,w) \in \{1,\dots,l\}^3$ such that $s_i s'_w \neq 0$ and $s_j s'_w \neq 0$.
	\end{Lemma}
	\begin{IEEEproof}
		Recall that $\bm{u}=\bm{A}^T \bm{s'} + \bm{e'}$. It holds that
		\begin{equation}
		(\bm{A}^T \bm{s'})_i = \sum_{j=1}^l A_{ji} s'_j \enspace .
		\end{equation}
		If one of the polynomials $s'_j\neq 0$ for $j\in\{1,\dots,l\}$, the sum in the equation above is uniformly distributed on $\RQ$. This can be shown by considering that we have a uniformly distributed summand for each polynomial coefficient and the assertion follows in accordance with the security proof for a one-time pad. Since the $A_{ji}$ are uniformly distributed, the $\bm{\compnoiseu}$ is independent of $\bm{e}$, $e''$ and $\bm{s}$ and furthermore independent of $\bm{s'}$ and $\bm{e'}$ under the condition that $\bm{s'}\neq 0$.
		
		In a very similar fashion it follows for $\compnoisev$ that it is independent of $\bm{e'}$ and independent of $\bm{s},\bm{s'},\bm{e},e''$ if a pair $(i,j) \in \{1,\dots,l\}^2$ exists such that $s_i s'j \neq 0$ by examining
		\begin{equation}\label{eq:compnoisev}
		\bm{s}^T\bm{A}^T\bm{s'} = \sum_{i=1}^l \sum_{j=1}^l s_i A_{ji} s'_j \enspace .
		\end{equation}
		Since $s_i s'_j \neq 0$ there is a non-zero coefficient within this product, say the coefficient with index $p$. Notice that this coefficient is multiplied with different coefficients of $A_{ji}$ for each coefficient of the product $s_i A_{ji} s'_j$.
		
		Next we show the independence of the compression noise $\compnoisev$ with respect to the other noise terms. The existence of indices $i,j,w$ such that $s_i s'_w \neq 0$ and $s_j s'_w \neq 0$ implies that $\compnoisev$ is decoupled from $\bm{\compnoiseu}$ by the one-time pad property.
		This holds because uniformly distributed elements of different rows in $\bm{A}^T$ contribute to the sum in equation~\eqref{eq:compnoisev}.
	\end{IEEEproof}
	
	\begin{Corollary}\label{cor:compression}
		Let an RLWE based scheme with parameters $(n,q,k)$ and only ciphertext compression in $v$ be given. If it holds that $ss'\neq 0$, then the noise distribution can be computed by convolving the distributions of the difference nosie terms and the compression noise $\compnoisev$.
	\end{Corollary}
	\begin{IEEEproof}
		This statement can be proved similarly to the statement for the compression noise $\compnoisev$ of Lemma~\ref{lem:independence_noise}.
	\end{IEEEproof}
	
	The necessary conditions for Lemma~\ref{lem:independence_noise} and Corollary~\ref{cor:compression} hold with overwhelming probability for the parameter sets of Kyber and NewHope considered in this work. Therefore, we neglect the unlikely event that the conditions for Lemma~\ref{lem:independence_noise} or Corollary~\ref{cor:compression} do not hold in the following. The analysis for the validity of this statement can be found in the appendix.

	\subsection{Transforming the Public Key Encryption scheme into an IND-CCA2 secure KEM}\label{subsec:fujisaki}
	
	A common requirement for key-encapsulation mechanisms (KEMs) is IND-CCA2-security. As a reference for explanation of different security notions we refer to \cite{katz2014introduction}.
	In \cite{hofheinz2017modular} a modular analysis of the Fujisaki-Okamoto transform \cite{fujisaki1999secure} is presented which enables the transformation of an IND-CPA secure PKE scheme into an IND-CCA secure KEM. The authors also address the problem of obtaining and IND-CCA secure scheme from an IND-CPA secure one even if the decryption failure rate is non-zero. For the security level however the authors mention that a small decryption failure rate is still very important. For a security level equivalent to AES256 for NewHope1024 \cite{alkim2019newhope} the decryption failure rate is upper bounded by $2^{-216}$ and for Kyber1024 \cite{Kyber_supp} the decryption failure rate is upper bounded by $2^{-174}$.
	
	\section{The RLWE/MLWE channel with increased alphabet size}\label{sec:RLWE_channel}

	\subsection{Channel Model}
	
\begin{figure}[t]
		\begin{center}
			\scalebox{0.87}{
				\includegraphics{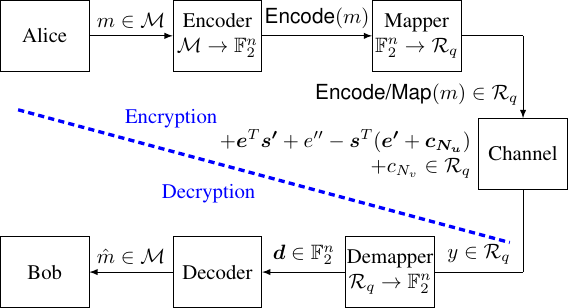}
				}
		\end{center}
		\caption{RLWE/MLWE channel}
		\label{fig:RLWE_channel}
		\vspace{-1.5em}
		
	\end{figure}

	The encryption and decryption procedure of RLWE-based and MLWE-based cryptosystems can be interpreted as the transmission of symbols over a communication channel with additive noise. The corresponding channel models which we call the RLWE channel and the MLWE channel are illustrated in Fig.~\ref{fig:RLWE_channel}. The sender's goal is to transmit a message $m$ contained in the message space $\mathcal{M}$ reliably to the receiver.
	
	The input to this channel as well as its output is a polynomial in $\RQ$.
	The additive noise on the channel follows the same distribution as the noise term within equation~\eqref{eq:decoding}. Notice that $s,e,s',e'$ are in $\RQ$ for the RLWE channel while $\bm{s,e,s',e'}$ are in $\RQ^l$ for the MLWE channel; $e''$ is in $\RQ$ in both cases.
	The crucial properties for the analysis to follow are the same for RLWE and MLWE based systems.
	
	Estimating DFRs for currently proposed schemes ($2^{-174}$ for Kyber, $2^{-216}$ for NewHope, cf.~Section~\ref{subsec:fujisaki}) using Monte Carlo simulations is infeasible.
	However, it is possible to compute the marginal distribution of the coefficient failure rate for one coefficient $\coefffailprob$. We will show how to obtain an upper bound on $\coefffailprob$ in Section~\ref{subsec:generalisation_Q_ary} (Theorem~\ref{th:bound_error_probability}).

	\subsection{Stochastic Independence Assumption}\label{subsec:independence_assumption}
	So far, it is unknown how to precisely obtain the $\decfailprob$ for RLWE/MLWE-based schemes.
	RLWE/MLWE channels have memory within each message block as the noise is generated from multiplying and adding several polynomials. The coefficients of these polynomials are therefore not stochastically independent due to the multiplications, implying that coefficient failures are not independent either.
	To estimate the $\decfailprob$, it is widely assumed that coefficient failures within a block occur independently with probability $\coefffailprob$ (cf.~\cite{Hila5_supp,fritzmann2018analysis,LAC_supp1}). The assumption of independent coefficient failures is not only common for lattice-based schemes but also for code-based schemes (e.g., HQC \cite{melchor2018hamming}, which is an alternative finalist in the NIST-PQC Round 3). Thus the RLWE/MLWE channel can be modelled as $n$ parallel \emph{Binary Symmetric Channels} with error probability $\coefffailprob$. To the best of our knowledge, due to the algebraic operations that connect the different components of the noise in a complicated manner there is no tight bound on the $\decfailprob$ that fully covers the dependency of the coefficient failures. Known upper bounds on the $\decfailprob$ not using this independence assumption are rather loose even though for NewHope the attempt presented in \cite{schanck2020upper} is worth mentioning but not applicable for the encoding schemes presented in this work.

	If algebraic codes with hard decision decoding are considered, the minimum distance $d$ determines the number of errors $t = \left\lfloor (d-1)/2 \right\rfloor$ up to which correct decoding can be guaranteed.
	Thus, if we consider stochastically independent coefficient failures an error-correcting code with error-correction capability of $t$ symbols leads to an overall failure rate of the scheme
	\begin{equation}\label{eq:failure_rate}
		\decfailprob \leq \sum_{j=t+1}^n \binom{n}{j} \coefffailprob^j (1-\coefffailprob)^{n-j} \enspace .
	\end{equation}
	
	In general however, the the assumption of independent coefficient failures does not hold in RLWE/MLWE-based schemes and it has been shown in \cite{d2019impact} that the stochastic dependence between coefficient failures has a significant impact on the $\decfailprob$ of LAC \cite{LAC_supp1}. Hence, the LAC team changed the error distribution for polynomials in their Round 2 submission for the NIST-PQC \cite{LAC_supp}.
	In \cite{maringer2019} it has been experimentally shown that this significantly reduces the stochastic dependence of coefficient failures. Quantitative statements have been obtained by using statistical methods.
	
	Although we know that estimating the DFR by using the assumption of independent coefficient failures is not completely accurate we consider it to be a reasonable first order approximation of the real behavior of RLWE/MLWE-based cryptosystems.
	We state clearly throughout this work whenever we make use of this assumption.

	\subsection{Generalization to $\alsize$-ary alphabets}\label{subsec:generalisation_Q_ary}
	
	It is natural to extend the channel input alphabet to be $\alsize$-ary.
	Consider the encoding procedure. In all of the discussed realizations of the RLWE/MLWE schemes, the ECC was binary and the message was mapped to a polynomial with coefficients in $\{0,\left \lfloor q/2 \right \rfloor\}$.
	We extend the channel to $\alsize$-ary alphabets by splitting $\left [-\left\lfloor q/2 \right\rfloor,\left\lfloor q/2 \right\rfloor \right ]$ into smaller intervals of size either $\left \lfloor q/\alsize \right\rfloor$ or $\left \lceil q/\alsize \right \rceil$, where their respective occurrence depends on the remainder of the integer division of $q$ by $\alsize$. This approach has already been followed for the LWE-based scheme Frodo in \cite{bos2016frodo}. Notice that since we are in $\mathbb{Z}_q$ distance is defined to be cyclic. Therefore, it is perfectly fine that $-\left\lfloor q/2 \right \rfloor$ and $\left\lfloor q/2 \right \rfloor$ belong to the same quantization interval as they only have a distance of $1$ for odd $q$. The appropriate distance measure for these kinds of considerations is the Lee-metric \cite{lee1958some}. It is defined as $d_L(x,y):=\min(|x-y|,q-|x-y|)$.
	
	The channel's input alphabet consists of the central elements of these intervals. The deployed ECC is changed to be of $\alsize$-ary alphabet size and the mapper's output alphabet is defined to be equal to the channel's input alphabet.
	
	A possible choice for the demapper is to extend the hard decision demapping procedure of LAC to $\alsize$-ary alphabet sizes. The demapping procedure within LAC can be interpreted as a linear quantization of $\left [ -\left\lfloor q/2 \right\rfloor, \left\lfloor q/2 \right\rfloor \right ]$.
	Recall the subintervals considered in the construction of the mapper.
	We define these subintervals to be the quantization intervals and their center points to be the respective reproduction values. By this methodology we have generalized $\encmap$ and $\demapdec$ for $\alsize=2$ to arbitrary $\alsize$.
	Basically the receiver uses the quantizer to estimate the symbols transmitted by the sender and uses the ECC to correct possibly erroneous symbols.
	We remark that the choice of the demapper is by no means optimal because soft information is not utilized.
	Notice that the difference of the decryption function for different alphabet sizes lies entirely in $\demapdec$ and in particular Equation~\eqref{eq:decoding} does not change if we consider the $\alsize$-ary case because all the information about the input alphabet size is contained in the functions $\encmap$ and $\demapdec$.

	The following theorem is based on a result in \cite{lindner2011better} and proves that generalizing RLWE/MLWE-based schemes to \\$\alsize \geq 2$ does not necessarily decrease their security level.

	\begin{Theorem}\label{th:increase_alsize}
		The security level of RLWE/MLWE-based schemes is not reduced by the generalization to a $\alsize$-ary alphabet as long as the decryption failure rate is not increased.
	\end{Theorem}
	\begin{IEEEproof}
		Due to Lemma~\ref{lem:indistinguishability_ciphertext} it holds that the tuple $(\bm{u},v)$ cannot be distinguished from a uniformly distributed sample on $\RQ^l \times \RQ$ under the MLWE assumption. By following exactly the same steps as in its proof, it follows that the value of $\encmap(m)$ has no influence on the distribution of $(\bm{u},v)$ irrespective of $\alsize$.
		As already mentioned in Subsection~\ref{subsec:fujisaki} a low decryption failure rate is essential to obtain a high security level for the resulting scheme after the transformation into an IND-CCA secure KEM. Hence, we have to avoid increasing the $\decfailprob$ in order to keep the same security level.
	\end{IEEEproof}
	
	Generalizing the RLWE/MLWE-based scheme to $\alsize$-ary input alphabets increases the coefficient failure probability if all other parameters of the system remain the same. To avoid increasing the $\decfailprob$, the error-correction capability of the deployed ECC has to be increased accordingly.
	
	We consider the generalization of the demapping strategy of LAC for the $\alsize$-ary case and we show how to upper bound the coefficient failure probability $\coefffailprob$ which can then be used to obtain an upper bound on the decryption failure rate similar to~\eqref{eq:failure_rate}.
	We define $\psi$ to be the probability distribution of the $i$-th coefficient of the  noise $(\bm{e}^T\bm{s'}-\bm{s}^T(\bm{e'}+\bm{\compnoiseu})+e''+\compnoisev)_i$. Indexing the distribution $\psi$ is unnecessary in both cases because all coefficients of the difference noise are distributed in the same way due to the symmetry of $\chi_k$.
	
	In order to compute an upper bound on $\coefffailprob$ for RLWE/MLWE channels we first prove the following Lemma.

	\begin{Lemma}[Noise distribution for MLWE]\label{lem:MLWE_psi}
		Recall that the noise is given by $\bm{e}^T\bm{s'} + e''- \bm{s}^T(\bm{e'} + \bm{\compnoiseu})+\compnoisev$. We define the distribution of the product of two elements in $\mathbb{Z}_q$ which have been sampled according to the error distribution $\chi_k$ by $\xi_k$. Furthermore, we define the distribution of one coefficient of $\bm{s}^T(\bm{e'} + \bm{\compnoiseu})$ by $\eta_k$ and the distribution of $\compnoisev$ by $\rho_v$. Then for the MLWE channel it holds
		\begin{equation}\label{eq:MLWE_psi}
			\psi = \oast_{l-1}(\oast_{n-1}(\xi_k))  * \eta_k * \chi_k * \rho_v
		\end{equation}
	\end{Lemma}
	\begin{IEEEproof}
		Consider the product of two polynomials $a, b \in \RQ$ sampled according to $\chi_k(\RQ)$. The $i$-th coefficient of their product equals
		\begin{equation}\label{eq:product_poly}
		(ab)_i = \sum_{j=0}^i a_j b_{i-j} - \sum_{j=i+1}^{n-1} a_j b_{n-j+i} \enspace .
		\end{equation}
		We remark that addition and subtraction of polynomials sampled according to $\xi_k$ or $\chi_k$ leads to the same resulting distributions due to the symmetry of the distribution $\chi_k$ around zero. Since the first summand of the noise is $\bm{e}^T\bm{s'}$ we obtain its distribution by first summing $n$ terms that are distributed according to $\xi_k$ for one polynomial multiplication and then summing $l$ terms that are distributed according to the resulting distribution to compute the scalar product.
		To obtain the overall noise the result is added to coefficient of $\bm{s}^T(\bm{e'} + \bm{\compnoiseu})$ which is distributed according to $\eta_k$, a coefficient of $e''$ which is distributed according to $\chi_k$ and a coefficient of $\compnoisev$ distributed according to $\rho_v$. The assertion follows from these considerations. Notice that the computation of the noise distribution makes use of Lemma~\ref{lem:independence_noise}.
	\end{IEEEproof}
	
	The computation of the distribution $\eta_k$ of a single coefficient of $\bm{s}^T(\bm{e'}+\bm{\compnoiseu})$ can be easily computed for the parameter set of Kyber that we consider in this work (Kyber1024). 
	\begin{Remark}
		Within NewHope $\psi = \oast_{n-1}(\xi_k) * \oast_{n-1}(\xi_k) * \chi_k * \rho_v$ because the first component of the ciphertext is not compressed, i.e. $\compnoiseu = 0$.
	\end{Remark}

	\begin{Theorem}\label{th:bound_error_probability}
		Let the alphabet size be $\alsize$ and let the probability distribution of a coefficient of the difference noise be denoted again by $\psi$. Then the length of every demapping (quantization) interval is at least $\lfloor q/\alsize\rfloor$ and it holds that
		
		\begin{equation}\label{eq:single_coeff_failure}
			\coefffailprob \leq 1 - \sum_{i=-\lfloor q/(2\alsize)\rfloor}^{\lfloor q/(2\alsize)\rfloor} \psi(i) =: \overline{\coefffailprob} \enspace .
		\end{equation}
		
		Assuming coefficient failures to occur stochastically independent with respect to each other it follows that
		\begin{equation}\label{eq:bound_failure_woe}
			\decfailprob \leq \sum_{j=t+1}^n \binom{n}{j} \overline{\coefffailprob}^j (1-\overline{\coefffailprob})^{n-j} \enspace .
		\end{equation}
	\end{Theorem}
	\begin{IEEEproof}
		We will show this statement by proving that the probability of a successful reception is lower bounded by
		
		\begin{equation}\label{eq:lower_bound_correct}
			\sum_{i=-\lfloor q/(2\alsize)\rfloor}^{\lfloor q/(2\alsize)\rfloor} \psi(i) \enspace .
		\end{equation}
		
		Indeed, if we choose the quantization intervals according to section~\ref{subsec:generalisation_Q_ary} and put the reconstruction values into the middle of the intervals we obtain that the probability for a correct symbol is lower bounded by \eqref{eq:lower_bound_correct}. 
		This statement implies the upper bound given in~\eqref{eq:single_coeff_failure}.
		Inequality~\eqref{eq:bound_failure_woe} follows from~\eqref{eq:single_coeff_failure} by using a standard combinatorial argument.
	\end{IEEEproof}
	
	\section{Information-Theoretic Analysis of the RLWE/MLWE channel}\label{sec:information_theoretic_LWE}

	\subsection{A Lower bound on the capacity of the RLWE/MLWE-channel}
	
	\begin{figure}
		\centering
		\scalebox{0.85}{
			\includegraphics{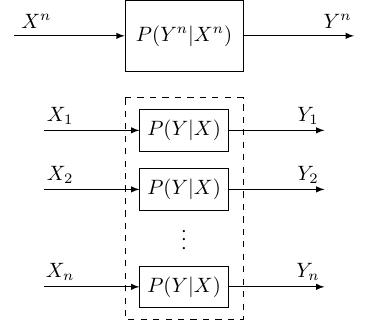}}
		\vspace{-0.5em}
		\caption{RLWE/MLWE and marginalized RLWE/MLWE channel (dashed)}
		\vspace{-1em}
		\label{fig:RLWE/MLWE both}
	\end{figure}

	We define an auxiliary channel consisting of $n$ parallel channels which are defined by the marginalization of $P(Y^n|X^n)$ for one coefficient. We label its distribution by $P(Y|X)$, where the distribution is independent of the index within the RLWE/MLWE block due to the symmetry of $\chi_k$. This auxiliary channel is in the following referred to as the \emph{marginalized RLWE/MLWE channel}.
	Fig.~\ref{fig:RLWE/MLWE both} depicts RLWE/MLWE channel and marginalized RLWE/MLWE channel.
	The channel inputs $X_1\ldots X_n$ denote the coefficients of the polynomial $\encmap(m)$ and the corresponding channel outputs $Y_1 \ldots Y_n$ denote the coefficients of $y \in \RQ$.
	
	\begin{Lemma}\label{lem:indep_channel_alternative}
		Let $X_i$ denote the $i$-th input symbol to the RLWE/MLWE channel and let $Y_i$ denote the $i$-th output symbol of the channel. We denote the vectors containing the sequences $X_1,\dots,X_k$ and $Y_1,\dots,Y_k$ by $X^k$ and $Y^k$, respectively.
		Let $H(X_i) = H(X_j)$ and let $H(X_i|Y_i) = H(X_j|Y_j) \; \forall i,j$. Furthermore, let the input symbols to the channel be stochastically independent, then
		\begin{equation}\label{eq:lem4}
		I(X^n;Y^n) \geq nI(X;Y)\enspace ,
		\end{equation}
		where we omitted the indices on the right hand side of the previous inequality because the mutual information between $I(X_i;Y_i)$ does not depend on the respective index $i$.
	\end{Lemma}
	\begin{IEEEproof}
		\begin{align}
		I(X^n;Y^n) &= H(X^n) - H(X^n|Y^n) \nonumber\\
		&= nH(X) - \sum_{i=1}^n H(X_i|Y^nX^{k-1})\nonumber\\
		&\geq nH(X) - \sum_{i=1}^n H(X_i|Y_i)\nonumber \\
		&= nH(X) - nH(X|Y) = nI(X;Y)
		\end{align}
	\end{IEEEproof}
	Notice that in the conditions of Lemma~\ref{lem:indep_channel_alternative} the input symbols are independent rather than the channel being memoryless. The statement of Lemma~\ref{lem:indep_channel_alternative} is therefore not to be confused with a standard result in information theory stating that the inequality in \eqref{eq:lem4} holds in the opposite direction for discrete memoryless channels (irrespective of the input distribution) \cite[Lemma 7.9.2]{cover1999elements}.
	
	Since the marginalized RLWE/MLWE channel is composed of $n$ identical component channels its capacity can be computed by determining the capacity of one component channel and multiplying the result by $n$.
	Let $X$ be a random variable modelling the input distribution of one component channel and $Y$ be the random variable specifying its output. We denote the ranges of $X$ and $Y$ by $\mathcal{X}$ and $\mathcal{Y}$, respectively, where $\lvert \mathcal{X} \rvert = \alsize$.
	
	\begin{Lemma}\label{lem:uni_disp}
		Each component channel of the marginalized RLWE/MLWE channel belongs to the class of uniformly dispersive channels, meaning that the set $\{P(y|x): y \in \mathcal{Y}\}$ is the same for all $x \in \mathcal{X}$ and it holds that
		\begin{equation}
			H(Y|X) = H(Y|X=x) = H(\psi)
		\end{equation}
		for all $x \in \text{supp} (P_X)$ where $\psi$ denotes the distribution of one coefficient according to the channel noise (Lemma~\ref{lem:MLWE_psi}).
	\end{Lemma}
	\begin{IEEEproof}
		By the definition of the conditional entropy we have
		\begin{equation}\label{eq:cond_entropy}
			H(Y|X) = \sum_{x \in \supp (P_X)} H(Y|X=x) P_X(x)
		\end{equation}
		Without loss of generality we assume that we analyze the $i$-th component channel and therefore its output \\$Y_i=(\encmap(m))_i + (\bm{e}^T\bm{s'}-\bm{s}^T(\bm{e'}+\bm{\compnoiseu})+e''+\compnoisev)_i$, where $(\encmap(m))_i$ denotes the $i-th$ component of the encoded message after the mapper.

		Since $X_i = (\encmap(m))_i$ it follows that\\
		$P_{Y|X=x} = \psi, \; \forall x \in \text{supp}(P_X)$
		and therefore \\$\{P(y|x): y \in \mathcal{Y}\}$ is the same $\forall x \in \text{supp}(P_X)$ which implies
			$H(Y|X=x) = H(\psi), \; \forall x \in \text{supp}(P_X)$.
		Applying this result to \eqref{eq:cond_entropy} proves the first equality of this lemma.
	\end{IEEEproof}

	The following corollary is standard textbook knowledge in the field of information theory and can for instance be found in \cite{massey1998applied}. We recap it here for the the sake of completeness.
	\begin{Lemma}\label{cor:uniformly_dispersive}
		For uniformly dispersive channels the channel capacity is equal to
		\begin{equation}
			C = \max_{P_X} H(Y) - H(Y|X=a)
		\end{equation}
		for some $a \in \supp(P_X)$.
	\end{Lemma}

	Lemma~\ref{cor:uniformly_dispersive} states that for uniformly dispersive channels the maximization of the mutual information boils down to the maximization of an entropy. Therefore, we aim at finding $P_X$ such that $H(Y)$ is maximized.
	We start by stating the following lemma which is given as an exercise in \cite{cover1999elements}.
	\begin{Lemma}\label{lem:entropy_monotonicity}
		Consider a random variable $Y$ with distribution $P_Y$ and consider a random variable $Z$ with the same distribution except for $k$ events, denoted as $a_{i_1},\dots,a_{i_k}$, where $P_Z(a_{i_1}) = \dots = P_Z(a_{i_k}) = \sum_{j=1}^k P_Y(a_{i_j})/k$.
		Then it holds that $H(Y) \leq H(Z)$.
	\end{Lemma}
	
	Next we make use of Lemma~\ref{lem:entropy_monotonicity} to show that the uniform distribution achieves capacity if $\alsize$ divides $q$.
	\begin{Theorem}\label{th:shaping}
		If $q$ is divisible by $\alsize$ the uniform distribution on $\mathcal{X}$ achieves the capacity of a component channel of the marginalized RLWE/MLWE channel.
		
		The distribution of the output $Y$ can then be computed by:
		\begin{equation}\label{eq:output_distribution}
			P_Y(y) = \frac{1}{\alsize} \sum_{j=0}^{\alsize-1} \psi \left(y+\frac{jq}{\alsize}\right)
		\end{equation}
	\end{Theorem}
	\begin{IEEEproof}
	We know due to Lemma~\ref{cor:uniformly_dispersive} that the problem of maximizing $I(X;Y)$ can be reduced to maximizing the entropy of the output distribution $H(Y)$. Since any finite dimensional cube $[0,1]^\alsize$ is compact and the distribution $H(Y)$ is a continuous function with respect to $P_X$ we know that there exists some $P_X^*$ maximizing $H(Y)$. Suppose that we are provided with this distribution. We take the resulting output distribution $P_Y^*$ and observe its values for the set $\{i,i+q/\alsize,i+2q/\alsize,\dots,i+(\alsize-1)q/\alsize\}$.
	\begin{align*}
	P_Y^*(i) &= P_X^*(0) \psi(i) + P_X^*(1) \psi \left( i-\tfrac{q}{\alsize}\right) + \dots \\
	&\quad + P_X^*(\alsize - 1) \psi \left( i-\tfrac{(\alsize-1)q}{\alsize} \right)\\
	P_Y^*\left(i+\tfrac{q}{\alsize}\right) &= P_X^*(0) \psi \left( i+\tfrac{q}{\alsize}\right) + P_X^*(1) \psi(i) + \dots \\
	&\quad + P_X^*\left( \alsize -1 \right) \psi \left( k-\tfrac{(\alsize -2)q}{\alsize}\right)\\
	&\hspace{1cm}\vdots \\
	P_Y^*\left( i+\tfrac{(\alsize-1)q}{\alsize}\right) &= P_X^*(0)\psi \left( i + \tfrac{(\alsize -1 )q}{\alsize}\right) +  \dots \\
	&\quad + P_X^*(\alsize - 1) \psi(i)
	\end{align*}
	Notice that the set of arguments within the function $\psi$ is the same for each equation. Furthermore, each possible pair of arguments with respect to $P_X^*$ and $P_Y^*$ occurs exactly once if all above equations are considered. Lemma~\ref{lem:entropy_monotonicity} states that for some specific $i$ we can only increase the entropy $H(Y)$ if we change $P_Y^*$ by replacing all values $P_Y^*(i),\dots,P_Y^*\left(i+(\alsize-1)q/\alsize\right)$ with their average without changing the remaining values of $P_Y^*$. This procedure can be performed for every $i$ without decreasing $H(Y)$ and is equivalent to changing $P_X^*$ to be the uniform distribution on the set $\mathcal{X}$ which concludes the proof.
	\end{IEEEproof}
	
	Using Lemma~\ref{lem:indep_channel_alternative} we give a lower bound on the channel capacity of the RLWE/MLWE channel.
	\begin{Theorem}\label{th:main_theorem}
		Let $n$ denote the RLWE/MLWE blocklength, let $X_i$ and $Y_i$ be the random variables specifying the $i$-th channel input and output, respectively, and let $P_X^{\otimes n}$ be the set of the product distributions on $X^n$, meaning that $X_1,\dots,X_n$ are independent identically distributed random variables according to some probability distribution $P_X$.
		Recall that $\psi=\oast_{l-1}(\oast_{n-1}(\xi_k))  * \eta_k * \chi_k * \rho_v$ denotes the distribution of one coefficient of the noise for the RLWE/MLWE channel (Lemma~\ref{lem:MLWE_psi}).
		
		The capacity of the RLWE/MLWE channel $C_{RLWE/MLWE}$ is lower-bounded by
		\begin{align}\label{eq:C_RLWE_MLWE}
		C_{RLWE/MLWE}&= \max_{P_{X^n}} I(X^n;Y^n) \geq \max_{P_X^{\otimes n}} I(X^n;Y^n) \nonumber \\
		&\geq \max_{P_{X}^{\otimes n}} \sum_{i=1}^n I(X_i;Y_i) = n \max_{P_{X}} I(X;Y) \enspace .
		\end{align}
		Furthermore, it holds that
		\begin{equation}\label{eq:bound_RLWE}
		C_{RLWE/MLWE} \geq n \left( H\left(\frac{1}{\alsize} \sum_{j=0}^{\alsize-1} \psi_j\right)  - H(\psi) \right) \enspace ,
		\end{equation}
		where we define $\psi_j(x):=\psi \left(x+\left \lfloor jq/\alsize\right\rfloor \right)$.
	\end{Theorem}
	\begin{IEEEproof}
		The channel coding theorem states that the capacity of a DMC with input $U$ and output $V$ is equal to $C = \max_{P_U} I(U;V)$. 
		In the context of the RLWE-channel $U \widehat{=} X^n$ and $V \widehat{=} Y^n$ and therefore the first equality in \eqref{eq:C_RLWE_MLWE} follows. The subsequent inequality follows due to the restriction of the maximization from the set of all distributions on $X^n$ to the set of product distributions. Furthermore, as a consequence the requirements for applying Lemma~\ref{lem:indep_channel_alternative} are fulfilled because the inputs are independently and identically distributed and $H(X_i|Y_i) = H(X_j|Y_j) \; \forall i,j$ due to the symmetry of the error distribution. This shows the second inequality in \eqref{eq:C_RLWE_MLWE}.
		
		Recall that by Lemma~\ref{cor:uniformly_dispersive} the last maximization in \eqref{eq:C_RLWE_MLWE} reduces to a maximization of $H(Y)$.
		Choosing a particular distribution at the input, e.g. the uniform distribution on the set $\mathcal{X}$, leads to a lower bound the channel capacity.
		For this input distribution it holds that 
		$$I(X;Y) = H(Y) - H(Y|X) = H\left(\frac{1}{\alsize} \sum_{j=0}^{\alsize-1} \psi_j\right) - H(\psi)\enspace ,$$
		where the entropy of the output distribution can be computed in accordance with the proof of Theorem~\ref{th:shaping} and the conditional entropy is given by Lemma~\ref{lem:uni_disp}.
	\end{IEEEproof}
	We emphasize that Theorem~\ref{th:main_theorem} shows that the capacity of the RLWE/MLWE channel is lower bounded by the capacity of the marginalized RLWE/MLWE channel which is equal to $\max_{P_{X}^{\otimes n}} \sum_{i=1}^n I(X_i;Y_i)$. 
	\begin{Remark}
		The difference between the optimized mutual information $\max_{P_X}I(X;Y)$ and $I(X;Y)$ for $P_X$ being chosen to be uniform has been experimentally observed not to be significant for the parameter sets considered in this work.
	\end{Remark}
	
	\subsection{The quantized RLWE/MLWE channel}
	The demapper implemented in LAC outputs each symbol by making a hard decision. By generalizing this demapper to the $\alsize$-ary case as described in Section~\ref{subsec:generalisation_Q_ary}, it is possible to define a new channel which we refer to as the \emph{quantized RLWE/MLWE channel}. This channel encapsulates the mapper, the RLWE/MLWE channel and the demapper into one channel which can be analyzed similarly to the RLWE/MLWE channel.
	
	In contrast to the RLWE/MLWE channel, soft information can only be used during the decoding process rather than throughout the combined process of demapping and decoding. 
	Therefore, the capacity of the quantized RLWE/MLWE channel is lower than the capacity of the RLWE/MLWE channel.
	Obtaining a lower bound on the capacity of the quantized RLWE/MLWE channel is very similar to the statements in Theorem~\ref{th:main_theorem}. The fixed demapper just quantizes the output of the RLWE/MLWE channel $Y$ which is distributed according to the probabilities of the quantization intervals. Notice that this channel is not uniformly dispersive if $\alsize$ does not divide $q$ but the lower bound in \eqref{eq:C_RLWE_MLWE} is still valid.

	\subsection{Plaintext bits per ciphertext bit}
	So far we have shown how to obtain lower bounds on the capacities for the RLWE/MLWE channels and their quantized counterparts.
	In practical terms it may however be more important how many plaintextbits can be transmitted over the channel per ciphertextbit. A scheme corresponding to a channel with high capacity does not necessarily perform well in terms of plaintextbits per ciphertextbit. This effect occurs because the parameter $l$ as well as the deployed ciphertext compression play an important role for this figure of merit.
	
	Lower bounds on the maximal amount of plaintextbits per ciphertextbit can be computed for both NewHope and Kyber from the lower bounds on the capacities determined by the methodology described in Theorem~\ref{th:main_theorem}.
	
	\begin{Proposition}\label{prop:plain/cipher}
		Let an MLWE based encryption scheme with ciphertext compression parameters $d_u$ and $d_v$ be given.
		Then the amount of plaintextbits per ciphertextbit for a given bitrate $R$ is given by the following formula:
		\begin{equation*}
		\frac{R}{l d_u + d_v}
		\end{equation*}
	\end{Proposition}
	\begin{IEEEproof}
		The amount of plaintextbits for a ciphertextblocklength of $n$ symbols is equal to $R \cdot n$. By the definition of the ciphertext compression we have that the corresponding ciphertext is of size $n(l \cdot d_u + d_v)$ bits. Dividing the amount of plaintextbits by the size of the corresponding ciphertext we obtain the statement of the proposition.
	\end{IEEEproof}

	\subsection{Results for NewHope and Kyber}\label{subsec:influence_channel_capacity}
	This subsection shows the examines the influence of $\alsize$ on NewHope with $(n,q,k,d_u,d_v)=(1024,12289,8,0,3)$ and Kyber with $(n,q,k,l,d_u,d_v) = (256,3329,2,4,11,5)$.
	
\begin{figure}[t!]
		\centering
		\scalebox{0.9}{
			\includegraphics{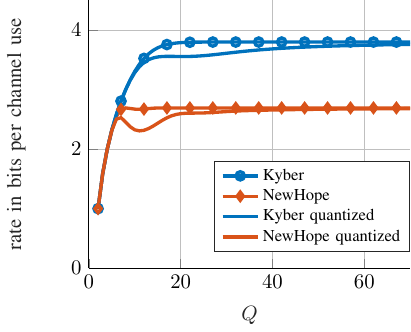}}
		\vspace{-1em}
		\caption{Lower bounds on the capacities of the RLWE/MLWE channels according to NewHope and Kyber including ciphertext compression}
		\label{fig:compressed_Mut_infs}
	\end{figure}

	In Fig.~\ref{fig:compressed_Mut_infs} the lower bounds on the capacity obtained by applying Theorem~\ref{th:main_theorem} to the aforementioned parameter sets for NewHope and Kyber are plotted for the RLWE/MLWE channel and the quantized RLWE/MLWE channel. For all parameter sets the results show that the lower bounds on the respective channel capacities can be significantly increased if larger input alphabet sizes $\alsize$ are considered compared to the originally proposed binary case.
	The plots also show that $\alsize$ only influences the lower bounds on the capacities for small alphabet sizes. The reason for this is that already for moderate alphabet sizes $\alsize$, $P_Y(y)$ is almost uniform such that increasing $\alsize$ further cannot significantly increase $H(Y)$ and therefore the capacity of the RLWE/MLWE channels since it is uniformly dispersive and therefore $H(Y|X)$ does not depend on $\alsize$. In the quantized case a similar effect occurs even though the alphabet size needs to be increased a bit in order to reduce the penalty on the achievable rate due to fixing the demapping strategy. In fact the intervals of the demapper are shrinking as $\alsize$ is increased until eventually each interval only contains one element and quantized RLWE/MLWE channel and RLWE/MLWE channel coincide.
	
\begin{figure}[t!]
		\centering
		\scalebox{0.9}{
			\includegraphics{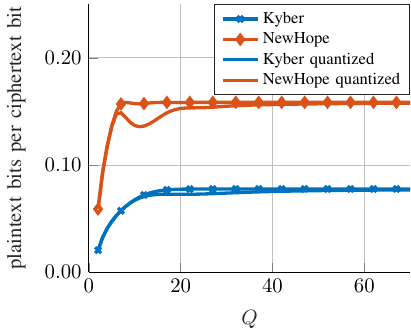}}
		\vspace{-1em}
		\caption{Lower bounds on the maximal amount of plaintextbits per ciphertextbits for the RLWE/MLWE channels of NewHope and Kyber}
		\label{fig:compressed_plain_cipher}
		\vspace{-1em}
	\end{figure}

	By applying Proposition~\ref{prop:plain/cipher} on the lower bounds on the capacity of the RLWE/MLWE channel and the quantized RLWE/MLWE channel, we obtain lower bounds on the maximal amount of plaintextbits per ciphertextbit for the respective channels. The results are presented in Fig.~\ref{fig:compressed_plain_cipher} and show that the lower bounds on the maximal amount of plaintextbits per ciphertextbit are higher for NewHope than for Kyber.
	
	\section{A semi-constructive analysis for obtainable rates for finite blocklengths and bounded decryption failure rates} \label{sec:semi_constructive_analysis}

	\subsection{Maximizing the achievable rate under $\decfailprob$ constraints} \label{subsec:maximizing_rate}
	Our goal in this section is to maximize the achievable rate for a single RLWE/MLWE block by varying $\alsize$ using error correcting codes to achieve the required $\decfailprob$s for NewHope and Kyber under the assumption of stochastically independent coefficient failures.
	We use Theorem~\ref{th:bound_error_probability} to find the required minimum distances $d$ for  $\alsize$-ary codes that guarantee the required decryption failure rates given in the supporting documentations of NewHope (less than $2^{-216}$) \cite{alkim2019newhope} and Kyber (less than $2^{-174}$) \cite{Kyber_supp}.
	We use the Gilbert--Varshamov bound to show that there exists a $\alsize$-ary linear code with minimum distance $d$ and dimension $k_{GV}$ for a given RLWE/MLWE blocklength $n$. We compute the resulting bitrate according to
	\begin{equation}\label{eq:GV_bitrate}
		R_{GV} := \frac{k_{GV}}{n} \log_2(\alsize) \enspace .
	\end{equation}
	Notice however that the Gilbert--Varshamov bounds only states that codes with parameters $[n, k_{GV}, d]_\alsize$ exist. In a second step, we therefore examine BCH codes of length at most $n$ achieving at least the required minimum distance $d$. These codes can be explicitly constructed and efficiently encoded and decoded.
	There are constraints on the length of BCH codes depending on their field size, e.g., there are no even length binary BCH codes. Since we cannot reduce the length of one RLWE block $n$ as this would have a negative effect on the security level of the scheme, we define the rates of BCH codes with respect to $n$, i.e.
	\begin{equation}\label{eq:BCH_bitrate}
		R_{BCH}:=\frac{k_{BCH}}{n} \log_2(\alsize) \enspace .
	\end{equation}

	For the parameter set $(n,q,k)=(1024,12289,8)$ of NewHope, Table~\ref{tab:NewHope_compression_final} shows that $\alsize = 4$ is optimal and that there is a BCH code achieving a bitrate $R_{BCH} = 1.7813$ at a $\decfailprob<2^{-216}$. Table~\ref{tab:Kyber_final} shows that for Kyber the optimal alphabet size is $\alsize=5$ and the best BCH code achieves a bitrate of $1.8412$ for $\decfailprob<2^{-174}$. For NewHope the original proposal achieves a rate of $0.25$ whereas for Kyber the original proposal does not deploy any ECC and therefore its bitrate equals $1$. The results for the best binary BCH code are omitted because an ECC is not necessary to achieve the required $\decfailprob$. However, for larger alphabet sizes higher rates can be achieved by using BCH codes. The results for all aforementioned schemes show that a substantial increase in bit rate is achievable by increasing $\alsize$ and using suitable ECCs.
	
	The capacity of the quantized RLWE/MLWE channel is an upper bound on the achievable rate by the concrete BCH code constructions. For NewHope and $\alsize=4$ there is a gap of about $0.2$ bit per channel use between the lower bound on the capacity and the rate achieved by the BCH code. Similarly this gap is about $0.5$ bit per channel use for Kyber for $\alsize=5$.
	
	\begin{table}[t!]
		\caption{Code parameters for different $\alsize$ for NewHope with $\decfailprob<2^{-216}$, $(n,q,k) = (1024,12289,8)$}
		\vspace{-1em}
		\begin{center}
			\setlength\tabcolsep{3pt}
			\begin{tabular}{ c c c c c c c c }
				$\alsize$ & $d$ & $k_{GV}$ & $R_{GV}$ & $n_{BCH}$ & $k_{BCH}$ & $R_{BCH}$ & plain/cipher\\
				\hline  
				$2$ & $3$ & $1014$ & $0.9902$ & $1023$ & $1013$ & $0.9893$ & $0.0582$\\
				$3$ & $11$ & $973$ & $1.5060$ & $1022$ & $949$ & $1.4689$ & $0.0864$\\
				$4$ & $31$ & $907$ & $1.7715$ & $1023$ & $912$ & $1.7813$ & $0.1048$\\
				$5$ & $81$ & $784$ & $1.7777$ & $939$ & $554$ & $1.2562$ & $0.0739$\\
				$7$ & $369$ & $344$ & $0.9431$ & $960$ & $91$ & $0.2495$ & $0.0147$\\
			\end{tabular}	
		\end{center}
		\label{tab:NewHope_compression_final}
		\vspace{-1.5em}
	\end{table}
	
	\begin{table}[t!]
		\caption{Code parameters for different $\alsize$ for Kyber with $\decfailprob<2^{-174}$, $(n,q,k,l) = (256,3329,2,4)$}
		\vspace{-1em}
		\setlength\tabcolsep{3pt}
		\begin{center}
			\begin{tabular}{ c c c c c c c c }
				$\alsize$ & $d$ & $k_{GV}$ & $R_{GV}$ & $n_{BCH}$ & $k_{BCH}$ & $R_{BCH}$ & plain/cipher\\
				\hline
				$2$ & $1$ & $256$ & $1$ & $-$ & $-$ & $-$ & $0.0204$\\
				$3$ & $5$ & $240$ & $1.4859$ & $242$ & $231$ & $1.4302$ & $0.0292$ \\
				$4$ & $9$ & $228$ & $1.7813$ & $255$ & $231$ & $1.8047$ & $0.0368$\\
				$5$ & $15$ & $214$ & $1.9410$ & $252$ & $203$ & $1.8412$ & $0.0376$\\
				$7$ & $33$ & $180$ & $1.9739$ & $240$ & $143$ & $1.5682$ & $0.0320$\\
			\end{tabular}	
		\end{center}
		\label{tab:Kyber_final}
		\vspace{-2em}
	\end{table}
	\subsection{Minimizing the $\decfailprob$ for a given minimum rate}	
		Designing NewHope in a way that the bit rate equals $0.25$ makes sense because it enables the transmission of $256$ bit of information within one RLWE block with $n=1024$. This corresponds to one AES256 key \cite{daemen1998block}. Similarly for Kyber the achieved rate has to be at least $1$ because for this scheme $n=256$.
		Public key algorithms are often used to share the key of a symmetric cryptosystem because those can be implemented very efficiently in hardware and symmetric algorithms are usually able to perform encryption quicker and without any ciphertext expansion. Due to Grover's algorithm~\cite{grover1996fast} the brute-force search of the key can be done in $\mathcal{O}(2^N/2)$ where $N$ denotes the length of the key in bit. Therefore, in order to obtain a post-quantum security level of $128$ bit a key having a length of at least $N=256$ bit is required for AES.
		
		However, if the the public key encryption (PKE) scheme shall directly be used to encrypt data or if in the future another symmetric cryptosystem with longer key size is used, it is sensible to transmit more data per ciphertext block. This can be useful if one would like to avoid an extra AES implementation to save chip area. Additionally, it is possible to share longer symmetric keys if that is necessary in the future using the same PKE system parameters.
		
		Table~\ref{tab:NewHope_Kyber_DFR} shows the largest minimum distances for NewHope and Kyber that achieve the required BCH-bitrates of $0.25$ and $1$, respectively, for different alphabet sizes $\alsize$. Furthermore the resulting $\decfailprob$s of the schemes under the assumption of independent coefficient failures are given. We observe that for NewHope $\alsize=2$ gives the lowest $\decfailprob = 2^{-12769}$ whereas for Kyber $\alsize=3$ is optimal resulting in $\decfailprob = 2^{-989}$. Notice that the optimal alphabet sizes for minimizing the $\decfailprob$ are different from the optimal alphabet sizes for maximizing the achievable rates for the required $\decfailprob$s in NewHope and Kyber.
		
	\begin{table}[t!]
		\caption{Decryption failure rates for different alphabet sizes $\alsize$ for a BCH-bitrate of at least $0.25$ for NewHope and $1$ for Kyber}
		\vspace{-1em}
		\begin{center}
			\setlength\tabcolsep{3pt}
			\begin{tabular}{ c c c c c }
				$\alsize$ & $d_{NewHope}$ & $d_{Kyber}$ & NewHope & Kyber\\
				\hline
				$2$ & $214$ & $1$ & $2^{-12769}$ & $2^{-174}$\\
				$3$ & $213$ & $26$ & $2^{-4307}$ & $2^{-989}$\\
				$4$ & $424$ & $46$ & $2^{-3646}$ & $2^{-953}$\\
				$5$ & $299$ & $44$ & $2^{-1075}$ & $2^{-547}$\\
				$7$ & $366$ & $59$ & $2^{-213}$ & $2^{-338}$\\
			\end{tabular}	
		\end{center}
		\label{tab:NewHope_Kyber_DFR}
		\vspace{-2em}
	\end{table}

	\subsection{Coding over multiple ciphertext blocks}
	In order to achieve capacity it is in general necessary to perform coding over infinitely long blocks. This is of course not possible but it was shown in \cite{arikan2009channel} that polar codes are capacity-achieving for the class of binary memoryless symmetric (BMS) channels. Furthermore, it has been shown in \cite{kudekar2017reed} that extended primitive narrow-sense BCH codes are capacity-achieving on the binary erasure channel (BEC) for blockwise MAP decoding.
	Even though the channel that we analyze in this work is no BEC we are still using BCH codes to obtain results in the finite length regime. The reason for this is that the claimed DFRs within the schemes are essential for the claimed security levels. Capacity-achieving codes using iterative decoding approaches like LDPC or polar codes cannot be simulated down to the DFRs we require for the systems under consideration in this work (e.g. for Kyber $2^{-174}$). In particular, LDPC codes feature error floors and thus need to be simulated down to the required DFR which is not feasible for the required DFRs of Kyber or NewHope. For polar codes there exist bounds on the DFR \cite{rajagopalan2018wiretap}, \cite{csacsouglu2009polarization} and therefore those are more suitable for LWE/RLWE/MLWE based cryptosystems. Using those upper bounds on the DFR reduces the achievable rate though and the DFRs we are aiming at are very small. We do not investigate polar codes further throughout this work but this could be an interesting point for further research if the bounds are tight enough.
	
	It is possible to use BCH codes over more than one ciphertext block to increase the achievable rate bringing it closer to the channel's capacity.
	In this work we took the approach to perform coding over four RLWE/MLWE blocks. This choice is arbitrary and has no particular reason, rather we took it as an example. The results of this approach are presented within Tables~\ref{tab:Kyber_extended}~and~\ref{tab:NewHope_extended}. Compared to the results in Subsection~\ref{subsec:maximizing_rate} we observe that coding over multiple blocks increases the achievable rates significantly, especially for larger input alphabet sizes.
	Notably, for Kyber the highest achievable rate is increased from $1.7813$ for $\alsize = 4$ to $2.0754$ for $\alsize = 7$. Notice also that for all $\alsize$ the achievable rate is increased and thereby closer to the lower bound on the channel capacity shown in Fig.~\ref{fig:compressed_Mut_infs}. For the other presented schemes similar behavior is observed.

			\begin{table}[t!]
				\caption{Code parameters for different $\alsize$ for Kyber with $\decfailprob<2^{-174}$, $(n,q,s,l) = (256,3329,2,4)$ for coding over $4$ blocks}
				\vspace{-1em}
				\setlength\tabcolsep{3pt}
				\begin{center}
					\begin{tabular}{ c c c c c c c c }
						$\alsize$ & $d$ & $k_{GV}$ & $R_{GV}$ & $n_{BCH}$ & $k_{BCH}$ & $R_{BCH}$ & plain/cipher\\
						\hline
						$3$ & $5$ & $1004$ & $1.5540$ & $1022$ & $997$ & $1.5432$ & $0.0315$\\
						$4$ & $9$ & $989$ & $1.9316$ & $1023$ & $993$ & $1.9395$ & $0.0396$\\
						$5$ & $17$ & $963$ & $2.1836$ & $1008$ & $899$ & $2.0385$ & $0.0416$\\
						$7$ & $41$ & $904$ & $2.4784$ & $960$ & $757$ & $2.0754$ & $0.0424$ \\
						$8$ & $59$ & $866$ & $2.5371$ & $1023$ & $738$ & $2.1621$ & $0.0441$\\
						$9$ & $87$ & $811$ & $2.5106$ & $1022$ & $631$ & $1.9533$ & $0.0399$\\
						\end{tabular}	
						\end{center}
						\label{tab:Kyber_extended}
						\vspace{-2em}
						\end{table}
						
						\begin{table}[t!]
							\caption{Code parameters for different $\alsize$ for NewHope with $\decfailprob<2^{-216}$, $(n,q,k) = (1024,12289,8)$ for coding over $4$ blocks}
							\vspace{-1em}
							\setlength\tabcolsep{3pt}
							\begin{center}
								\begin{tabular}{ c c c c c c c c }
									$\alsize$ & $d$ & $k_{GV}$ & $R_{GV}$ & $n_{BCH}$ & $k_{BCH}$ & $R_{BCH}$ & plain/cipher\\
									\hline
									$2$ & $3$ & $4084$ & $0.9971$ & $4095$ & $4083$ & $0.9968$ & $0.0586$\\
									$3$ & $13$ & $4021$ & $1.5559$ & $4088$ & $3992$ & $1.9204$ & $0.1130$\\
									$4$ & $37$ & $3924$ & $1.9160$ & $4095$ & $3933$ & $1.9395$ & $0.1141$\\
									$5$ & $115$ & $3679$ & $2.0855$ & $4069$ & $3388$ & $1.9206$ & $0.1130$\\
									$7$ & $829$ & $2277$ & $1.5606$ & $3268$ & $654$ & $0.4482$ & $0.0264$ \\
								\end{tabular}	
							\end{center}
							\label{tab:NewHope_extended}
							\vspace{-2em}
						\end{table}

	\section{Conclusion}\label{sec:conclusion}
	In this work we have shown how to treat RLWE/MLWE-based cryptosystems as communication channels. We have derived lower bounds on the channel capacities for the parameter sets of NewHope and Kyber for their highest proposed security levels. Our results show that enhancing the alphabet size of the channel input $\alsize$ increases the established lower bound on the channel capacity. Furthermore, we have shown why this effect saturates at a certain point. We proved why increasing $\alsize$ does not have a negative effect on the security level of RLWE/MLWE based cryptosystems as long as increasing the $\decfailprob$ is avoided. We have performed the same analysis for the quantized RLWE/MLWE channel.
	
	Under the assumption of stochastically independent coefficient failures we have presented achievability results regarding the bitrate based on the Gilbert-Varshamov bound for the parameter sets and required decryption failure rates of NewHope and Kyber. Recall that this bound does not give practical code constructions. Therefore, we have also given bitrates that can be achieved by using practically implementable BCH codes for the same paramter sets.
	Our results show that we are able to increase the bitrate of NewHope approximately by a factor of $7$ and that the rate of Kyber can be increased by a factor of $1.84$. Furthermore, we have shown that we can significantly reduce the decryption failure rates for NewHope and Kyber for fixed minimal bitrates of $0.25$ and $1$, respectively.
	
	\bibliographystyle{IEEEtran}
	\bibliography{mybibliography}
	
	\appendix
	\subsection{Conditions of Corollary~\ref{cor:compression} for NewHope}\label{subsec:conditions_NewHope}
	NewHope is specified not to have compression of the ciphertext component $\bm{u}$. Therefore, in this section we only consider compression of $v$ and its respective compression noise $\compnoisev$. This corresponds to the case described in Corollary~\ref{cor:compression}.
	
	We distinguish two ways in which $ss'$ can become equal to the zero polynomial, i.e. possibilities such that the Corollary is not applicable. The first one is that either $s$ or $s'$ is equal to zero. Since the error distribution $\chi_k$ is known, it is easy write a script computing that this probability is about $2.72 \cdot 10^{-724}$.
	The other possibility is that $ss' = (x^n+1) h$, where $h$ denotes an arbitrary non-zero polynomial in $\RQ$. It can be shown that for the parameter set $(n,q,k) = (1024,12289,8)$ the polynomial $x^n+1$ factorizes in linear factors with each linear factor occurring at most once by using a simple sage script. The zeros of the polynomial are not concentrated in a small subinterval of $[0,q-1]$ but rather distributed over the entire interval.
	
	It holds that
	\begin{equation*}
	ss' \equiv 0 \mod (x^n+1) \Leftrightarrow ss' = (x^n+1) h
	\end{equation*}
	for some polynomial $h \in \RQ$.
	In order for this scenario to occur all roots of $(x^n+1)$ have to occur at least once in either $s$ or $s'$. In the following we denote the roots of $(x^n+1)$ by $\alpha_1,\dots,\alpha_n$ and consequently
	\begin{align}
	s(x) &= (x-\alpha_{\pi(1)}) (x-\alpha_{\pi(2)}) \dots (x-\alpha_{\pi(j)}) \;a(x) \label{eq:a(x)}\\
	s'(x) &= (x-\alpha_{\pi(j+1)}) \dots (x-\alpha_{\pi(n)}) \;b(x) \label{eq:b(x)}
	\end{align}
	for some $j \in \{1,\dots,n-1\}$ and an arbitrary permutation $\pi$ of the set $\{1,\dots,n\}$.

	In contrast to the previous case where we computed the probability that one of the polynomials $s$ or $s'$ is equal to zero for the following analysis we consider them to be sampled from the uniform distribution on $\RQ$. This simplifies our analysis and we justify this methodology by the fact that the roots of $x^n+1$ are distributed over the entire interval $[0,q-1]$. Therefore, we assume it even to be more likely that within $s$ and $s'$ all roots $\alpha_1,\dots,\alpha_n$ are contained compared to the case where $s$ and $s'$ are sampled form $\chi_k$.
	
	Due to the uniform sampling of $s$ and $s'$ we can compute the probability that $ss'$ contains all roots by counting the pairs $(s,s')\in \RQ^2$ fulfilling this constraint and dividing this number by the total number of polynomials in $\RQ^2$ which is $q^{2n}$.
	
	\begin{Proposition}\label{prop:upper_NewHope}
		The probability that $ss'= (x^n+1) h$, for some non-zero polynomial $h \in \RQ$ is upper bounded by $\left(2/q\right)^n$.
	\end{Proposition}
	\begin{IEEEproof}
		The polynomials $x^n+1$ has $n$ roots which are to be distributed to the polynomials $s$ and $s'$.
		There are $\binom{n}{j}$ ways to choose $j$ roots of $x^n+1$ to be roots of $s$ whereas the remaining $n-j$ roots are to be roots of $s'$. In that case the polynomial $a(x)$ in \eqref{eq:a(x)} can be chosen arbitrarily from the set of polynomials with degree less than $n-j$. Therefore, there are $q^{n-j}$ possibilities for $s$. A similar argument shows that there are $q^j$ possibilities for $s'$. Summing over all possibilities of $j$ we obtain
		\begin{equation}
		\sum_{j=1}^{n-1} \binom{n}{j} q^{n-j} q^j = q^n(2^n-2)
		\end{equation}
		where we overcounted for instance the cases where $\alpha_1$ is a root in both $s$ and $s'$. Therefore, we get an upper bound on the number of possibilities for $s$ and $s'$ such that $ss'= (x^n+1) h$.
		Dividing this number by the amount of polynomials in $\RQ^2$ we obtain the desired result.
	\end{IEEEproof}
	By plugging the parameter set of NewHope into the upper bound of Proposition~\ref{prop:upper_NewHope} we obtain a value of about $3.9\cdot 10^{-3880}$. We consider the probabilities of both cases for which $ss'=0$ to be  small enough to consider the conditions of Corollary~\ref{cor:compression} to be fulfilled for the investigated parameter set of NewHope.

	\subsection{Conditions of Lemma~\ref{lem:independence_noise} for Kyber}\label{subsec:conditions_Kyber}
	Recall that for Kyber we have the parameter set $(n,q,k,l) = (256,3329,2,4)$. For this parameter set the polynomial $(x^n+1)$ factors into irreducible polynomials of order $2$. To guarantee independence of $\compnoisev$ from $\bm{\compnoiseu}$ and the terms generated by the difference noise combined, we require that there exists a pair $(i,j) \in \{1,\dots,l\}^2$ such that $s_i s'_j \neq 0$.
	
	Throughout this section we fix the indices $i,j,w$. We aim at showing that for these indices indeed the probability that the conditions of  Lemma~\ref{lem:independence_noise} are not fulfilled is very small.
	
	Similar to Subsection~\ref{subsec:conditions_NewHope} we split our analysis into two cases. For the first case we compute the probability that one of the polynomials $s_i,s'_j$ or $s'_w$ is equal to the zero polynomial. Again this computation can be performed by a simple script which shows that this probability is about $2^{-360}$ which is way below the desired security level of $256$ bit.
	
	The other possibility is that either $s_i s'_j = (x^n+1) h_1$ or $s_i s'_w = (x^n+1) h_2$, where $h_1$ and $h_2$ denote arbitrary non-zero polynomials in $\RQ$. It can be shown for the parameter set $(n,q,k,l) = (256,3329,2,4)$ that the polynomial $x^n+1$ factors into distinct irreducible polynomials of the form $x^2-\alpha_r$ with $\alpha_r$ being an element in $\mathbb{Z}_q$. Again this can be checked by using a simple sage script.
	
	\begin{Proposition}\label{prop:upper_Kyber}
		The probability that either $s_i s'_j = (x^n+1) h_1$ or $s_i s'_w=(x^n+1) h_2$, where both $h_1$ and $h_2$ are non-zero polynomials in $\RQ$ is upper bounded by $2^{n/2+1}/q^n$.
	\end{Proposition}
	\begin{IEEEproof}
		As in Subsection~\ref{subsec:conditions_NewHope} for the second step we consider the polynomials $s_i$, $s'_j$ and $s'_w$ to be sampled from the uniform distribution on $\RQ$. We aim at computing the probability that either $s_i s'_j = (x^n+1) h_1$ or $s_i s'_w = (x^n+1) h_2$, where $h_1$ and $h_2$ are non-zero elements in $\RQ$.
		By using the union bound and the fact that all elements are sampled independently from the same distribution we have that
		\begin{align*}
		Pr(s_i s'_j &= ((x^n+1) h_1 \lor s_i s'_w = (x^n+1) h_2) \\
		&\leq Pr(s_i s'_j = (x^n+1) h_1) + Pr(s_i s'_w = (x^n+1) h_2) \\
		&= 2 Pr(s_i s'_j = (x^n+1) h_1) \enspace .
		\end{align*}
		Therefore, we just need to show that the upper bound $Pr(s_i s'_j = (x^n+1) h_1) \leq 2^{n/2}/q^n$. 
		As in Subsection~\ref{subsec:conditions_NewHope} we investigate that we distribute the aforementioned irreducible polynomials to $s_i$ and $s'_j$ and upper bound the amount of pairs $(s_i,s'_j)$ fulfilling $s_i s'_j = (x^n+1) h_1$ for some non-zero polynomial $h_1$. The number of such pairs $(s_i,s'_j)$ is upper bounded (same overcounting argument as in Proposition~\ref{prop:upper_NewHope}) by
		\begin{equation*}
		\sum_{p=1}^{\frac{n}{2}-1} \binom{\frac{n}{2}}{p} q^{n-2p} q^{2p} = q^n (2^{n/2}-2)\leq q^n 2^{n/2} \enspace .
		\end{equation*}
		
	\end{IEEEproof}
	By plugging the parameter set of Kyber into the upper bound of Proposition~\ref{prop:upper_Kyber} we obtain a value of about $1.3\cdot 10^{-863}$. The probabilities for both analyzed cases are below the security level and therefore, we consider the conditions of Lemma~\ref{lem:independence_noise} to be fulfilled. Practically the likelihood that the required conditions for Lemma~\ref{lem:independence_noise} are not fulfilled are even much lower. We just avoided more complicated combinatorial arguments here.

	\subsection{Results for Frodo and LAC}

\begin{figure}[b!]
		\centering
		\scalebox{0.9}{
			\includegraphics{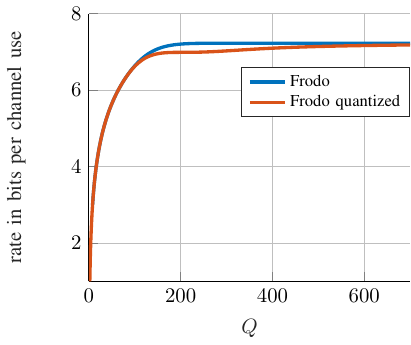}}
		\vspace{-1em}
		\caption{Lower bound on the capacity of the LWE channel according to Frodo}
		\label{fig:Frodo_Mut_infs}
	\end{figure}

\begin{figure}[b!]
	\centering
	\scalebox{0.9}{
		\includegraphics{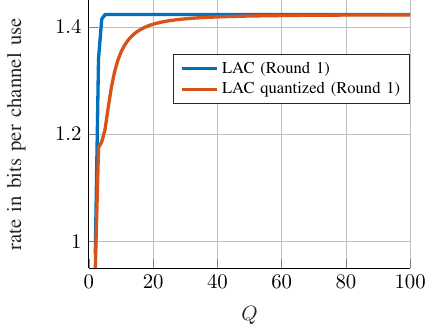}}
	\vspace{-1em}
	\caption{Lower bound on the capacity of the RLWE channel according to LAC (Round 1 submission)}
	\label{fig:LAC_Mut_infs}
	\end{figure}

\begin{figure}[t!]
		\centering
		\scalebox{0.9}{
			\includegraphics{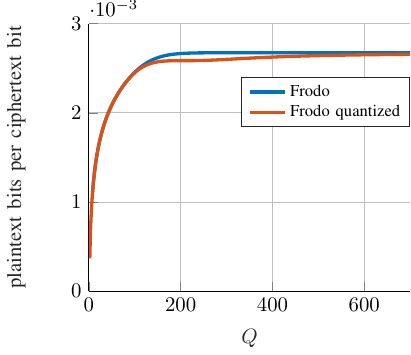}}
		\vspace{-1em}
		\caption{Lower bounds on the maximal amount of plaintextbits per ciphertextbits for the LWE channels of Frodo}
		\label{fig:Frodo_plaintextbits_ciphertextbits}
	\end{figure}

\begin{figure}[t!]
	\centering
	\scalebox{0.9}{
		\includegraphics{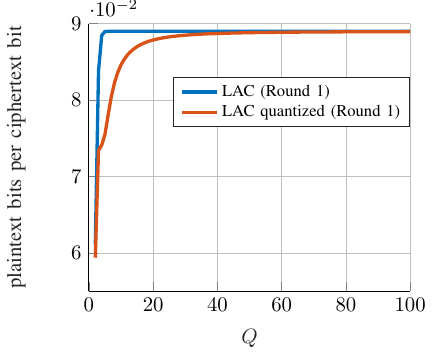}}
	\vspace{-1em}
	\caption{Lower bounds on the maximal amount of plaintextbits per ciphertextbits for the LWE channels of LAC (Round 1 submission)}
	\label{fig:LAC_plain_cipher}
	\end{figure}

	In this subsection we present the results for Frodo and LAC analogously to Sections~\ref{sec:information_theoretic_LWE}~and~\ref{sec:semi_constructive_analysis} using parameters for their highest respective security level. Most of the results for LAC have already been presented at the ITW 2020 \cite{maringer2020higher}. The results Frodo have not been published before and also show that our framework is applicable for standard LWE based schemes. The analysis for LWE based schemes is very similar to MLWE or RLWE based schemes and is therefore omitted. The applicability of the results to Frodo also demonstrates that the error distribution is not restricted to the centered binomial distribution but rather it can be adopted to other distributions (e.g. discrete Gaussian distributions) as well. Dedicated analysis for compression noise like we conducted in Appendixsubsection~\ref{subsec:conditions_NewHope} for NewHope and in Appendixsubsection~\ref{subsec:conditions_Kyber} is unnecessary for LAC and Frodo since within both schemes ciphertext compression is not deployed. For LAC we analyzed the initial submission for Round 1 of the NIST PQC competition because for our framework it is essential that the components of $s,e,s',e'$ need to be sampled independently which is not the case for the Round 2 submission, where the number of $+1$ and $-1$ elements within the error distribution are fixed. Fig.~\ref{fig:Frodo_Mut_infs} and Fig.~\ref{fig:LAC_Mut_infs} show the lower bounds on the achievable rates of Frodo and LAC, respectively. The respective bounds on the plaintext bits per ciphertext bits are presented in Fig.~\ref{fig:Frodo_plaintextbits_ciphertextbits} and Fig.~\ref{fig:LAC_plain_cipher}. The results for BCH coding over one ciphertext block coding are given in Tables~\ref{tab:Frodo_final}~and~\ref{tab:LAC} for Frodo and LAC, respectively. Furthermore, we give results for coding over $4$ LWE/RLWE blocks in Tables~\ref{tab:Frodo_extended}~and~\ref{tab:LAC_extended}.
	
		\begin{table}[b!]
		\caption{Code parameters for different $\alsize$ for Frodo with $\decfailprob<2^{-252.5}$, $(n,q,s,l) = (1,65536,6,8)$}
		\vspace{-1em}
		\setlength\tabcolsep{3pt}
		\begin{center}
			\begin{tabular}{ c c c c c c c c }
				$\alsize$ & $d$ & $k_{GV}$ & $R_{GV}$ & $n_{BCH}$ & $k_{BCH}$ & $R_{BCH}$ & plain/cipher\\
				\hline
				$16$ & $1$ & $-$ & $-$ & $-$ & $-$ & $-$ & $0.00148$\\
				$17$ & $3$ & $61$ & $3.8959$ & $64$ & $59$ & $3.7681$ & $0.00139$\\
				$19$ & $3$ & $61$ & $4.0488$ & $60$ & $57$ & $3.7833$ & $0.00140$\\
				$23$ & $5$ & $57$ & $4.0288$ & $63$ & $52$ & $3.6754$ & $0.00136$\\
				$27$ & $5$ & $57$ & $4.2348$ & $61$ & $51$ & $3.7891$ & $0.00140$ \\
				$29$ & $7$ & $54$ & $4.0989$ & $60$ & $51$ & $3.8712$ & $0.00143$\\
				$31$ & $7$ & $54$ & $4.1801$ & $64$ & $55$ & $4.2575$ & $0.00157$\\
				$32$ & $7$ & $54$ & $4.2188$ & $63$ & $45$ & $3.5156$ & $0.00130$\\
			\end{tabular}	
		\end{center}
		\label{tab:Frodo_final}
		\vspace{-1em}
	\end{table}

	\begin{table}[h]	
		\caption{Code parameters for different $\alsize$ for LAC256 and $\text{DFR}<2^{-115.4}$}
		\setlength\tabcolsep{3pt}
		\begin{tabular}{ c c c c c c c c c c c }
			$\alsize$ & $d$ & $k_{GV}$ & $R_{GV}$ & $n_{BCH}$ & $k_{BCH}$ & $R_{BCH}$ & plain/cipher\\
			\hline
			$2$ & $113$ & $521$ & $0.5088$ & $1023$ & $513$ & $0.5010$ & $0.0313$\\
			$3$ & $373$ & $183$ & $0.2833$ & $728$ & $51$ & $0.0789$ & $0.0049$\\
			$4$ & $659$ & $24$ & $0.0469$ & $1023$ & $32$ & $0.0469$ & $0.0029$\\
		\end{tabular}
		\label{tab:LAC}
		\vspace{-1em}
	\end{table}
	
	\begin{table}[b!]
		\caption{Code parameters for different $\alsize$ for Frodo with $\decfailprob<2^{-252.5}$, $(n,q,s,l) = (1,65536,6,8)$ for coding over $4$ blocks}
		\vspace{-1em}
		\setlength\tabcolsep{3pt}
		\begin{center}
			\begin{tabular}{ c c c c c c c c }
				$\alsize$ & $d$ & $k_{GV}$ & $R_{GV}$ & $n_{BCH}$ & $k_{BCH}$ & $R_{BCH}$ & plain/cipher\\
				\hline
				$19$ & $3$ & $253$ & $4.1981$ & $254$ & $250$ & $4.1484$ & $0.00153$\\
				$27$ & $5$ & $248$ & $4.6063$ & $242$ & $231$ & $4.2905$ & $0.00159$ \\
				$29$ & $7$ & $244$ & $4.6303$ & $240$ & $226$ & $4.2887$ & $0.00159$\\
				$31$ & $7$ & $244$ & $4.7220$ & $256$ & $232$ & $4.4897$ & $0.00166$\\
				$32$ & $7$ & $244$ & $4.7656$ & $255$ & $231$ & $4.5117$ & $0.00167$\\
				$47$ & $17$ & $226$ & $4.9037$ & $255$ & $211$ & $4.5782$ & $0.00169$\\
				$57$ & $27$ & $211$ & $4.8076$ & $250$ & $201$ & $4.5797$ & $0.00169$\\
				$67$ & $39$ & $194$ & $4.5970$ & $255$ & $143$ & $3.3885$ & $0.00125$
			\end{tabular}	
		\end{center}
		\label{tab:Frodo_extended}
		\vspace{-1em}
	\end{table}	

	\begin{table}[h]	
		\caption{Code parameters for different $\alsize$ for LAC256 and $\text{DFR}<2^{-115.4}$ for coding over $4$ blocks}
		\setlength\tabcolsep{3pt}
		\begin{tabular}{ c c c c c c c c c c }
			$\alsize$ & $d$ & $k_{GV}$ & $R_{GV}$ & $n_{BCH}$ & $k_{BCH}$ & $R_{BCH}$ & plain/cipher\\
			\hline
			$2$ & $211$ & $521$ & $0.5088$ & $4095$ & $2895$ & $0.7068$ & $0.0442$\\
		\end{tabular}
		\label{tab:LAC_extended}
		\vspace{-1em}
	\end{table}
\end{document}